\documentclass[reprint, amsmath, letter, amssymb, aps,  superscriptaddress, pra,]{revtex4-1}

\pdfoutput=1 
\usepackage[bookmarks=false, hidelinks]{hyperref}

\usepackage{graphicx}    
\usepackage{dcolumn}    
\usepackage{bm}           
\usepackage{float}
\usepackage{epstopdf}
\usepackage{dsfont}
\usepackage{bbold}

\usepackage[left=1.5cm, right=1.5cm, top=2.5cm, bottom=2.0cm]{geometry}

\newcommand{\bra}[1]{\langle{#1}\vert}
\newcommand{\ket}[1]{\vert{#1}\rangle}

\begin{document}
\title{Universal Set of Gates for Microwave Dressed-State \\ Quantum Computing}

\author{Gatis Mikelsons}
  \email{gmikelso@ic.ac.uk}
  \affiliation{Institute for Theoretical Physics, University of Ulm, Albert-Einstein-Allee 11, 89069 Ulm, Germany}
  \affiliation{Department of Physics, Imperial College London, SW7 2AZ, United Kingdom}
\author{Itsik Cohen}
  \affiliation{Racah Institute of Physics, The Hebrew University of Jerusalem, Jerusalem 91904, Givat Ram, Israel}
\author{Alex Retzker}
  \affiliation{Racah Institute of Physics, The Hebrew University of Jerusalem, Jerusalem 91904, Givat Ram, Israel}
\author{Martin B. Plenio}
  \affiliation{Institute for Theoretical Physics, University of Ulm, Albert-Einstein-Allee 11, 89069 Ulm, Germany}
  \affiliation{Department of Physics, Imperial College London, SW7 2AZ, United Kingdom}
\date{\today}
 
\begin{abstract}
We propose a set of techniques that enable universal quantum computing to be carried out using dressed states. This applies in particular to the effort of realising quantum computation in trapped ions using long-wavelength radiation, where coupling enhancement is achieved by means of static magnetic-field gradient. We show how the presence of dressing fields enables the construction of robust single and multi-qubit gates despite the unavoidable presence of magnetic noise, an approach that can be generalised to provide shielding in any analogous quantum system that relies on the coupling of electronic degrees of freedom via bosonic modes. 

\end{abstract}

\maketitle

\section{Introduction}

A promising experimental approach in the field of trapped-ion quantum information processing has been the introduction of microwave and radio wave sources. One particular technique, developed in the early 2000's, has involved making use of a static magnetic field gradient imposed along the trap axis to enhance particle interaction \cite{mintert2001ion}. This modification provides two crucial advantages. Firstly, by making the ions' equilibrium position dependent on the qubit state, the technique leads to much stronger coupling between motional and electronic states. This way, coupling to the ions' shared motional mode becomes possible even for long wavelength radiation, where the variability of the radiation field strength over the spatial extent of the ions' motional mode is effectively zero. This is quantised by the conventional Lamb-Dicke parameter, which is found to yield no useful interaction in the long-wavelength regime. However, it is found to be replaced by the effective Lamb-Dicke parameter in the static-gradient system \cite{mintert2001ion}, and this parameter is still large enough to enable useful quantum operations. Secondly, the presence of a magnetic gradient and the usage of magnetic-sensitive states spreads the resonance frequencies of the individual qubits, making them individually addressable even with long-wavelength radiation that is essentially impossible to focus in physical space to that resolution. Crucial building blocks of the scheme have been experimentally realised, notably sideband coupling \cite{johanning2008individual, HHH} and elements of conditional quantum logic \cite{PhysRevLett.108.220502}. 

A technique developed as an alternative to this approach has made use of oscillating magnetic fields inherent to near-field microwave radiation to realise elements of quantum dynamics using long-wavelength electromagnetic field \cite{ospelkaus2008trapped}. In this design, where ions are placed close to the microwave source, it is found that enough coupling between the motional and internal states of the ions becomes feasible. The issue of individual addressing is resolved by shifting the ions physically in space to alter the strength of the magnetic field experienced \cite{warring2013individual}. Implementation of microwave-driven single and multi-qubit gates using this route has been reported \cite{ospelkaus2011microwave}. 

In both designs, the usage of magnetic-sensitive states raises the issue of shielding the system from the unwanted effects of magnetic noise. In the oscillating-field design, atomic clock states are used, which are insensitive to magnetic field fluctuations to the first order. In the static-gradient case, a number of feasible strategies have been proposed. Pulsed decoupling \cite{viola1998dynamical, wokaun1987principles} provides one potentially useful approach. Alternatively, the usage of dressed states for encoding the logical qubit \cite{PhysRevA.62.042307, jonathan2001light, retzker2007fast, aharon2013general} offers a possible shielding technique. The dressed-state approach has previously found applications in resonator and nitrogen vacancy systems \cite{bermudez2011electron, PhysRevB.79.041302, cai2012long, cai2012robust, cai2013diamond} in addition to novel quantum gate designs for trapped ions using laser and laser-microwave addressing \cite{PhysRevA.85.040302, lemmer2013driven}. 

Notably, the dressed-state approach in the context of long-wavelength quantum computing with static magnetic gradients was explored by Timoney et al. in 2011 \cite{timoney2011quantum}, demonstrating experimentally its feasibility. Improvements in qubit coherence times by more than two orders of magnitude were reported. This exciting development holds the promise of robust, long-wavelength quantum computation, within the static-gradient approach, in a set-up that is experimentally viable and easily scalable.

Here, we address the next task of building a universal set of quantum gates for the microwave dressed-state approach in the static magnetic gradient set-up. Basic single-qubit operations for such a system have been realised by Timoney et al. \cite{timoney2011quantum} and also by Webster et al. \cite{webster2013simple} in a slightly modified arrangement.

We develop in detail the set-up employed in \cite{timoney2011quantum} and propose a set of quantum operations that jointly enable the execution of universal quantum computing. Firstly, we show how to realise arbitrary single-qubit rotations, proposing several alternative gate schemes. Secondly, following the well-known scheme of M{\o}lmer and S{\o}rensen \cite{MS, sorensen2000entanglement}, we develop a two-qubit entangling gate. We simulate the gates numerically to demonstrate their experimental viability and present analysis of the key noise sources. Finally, we comment on the possibilities for extending our scheme to the experimental set-up employed by Webster et al. \cite{webster2013simple}.

The techniques we develop are not directly transferable to the set-up where oscillating magnetic gradients are employed \cite{ospelkaus2008trapped}. Our work would suggest that even here states other than the clock qubits could be considered, in principle, replacing them with magnetically shielded dressed states. However, such an arrangement would result in a heavier experimental overhead and a reduction in the coherence times obtainable, as compared to the clock states.

\section{Physical system and definitions}

\begin{figure}
\includegraphics[width=0.38\textwidth]{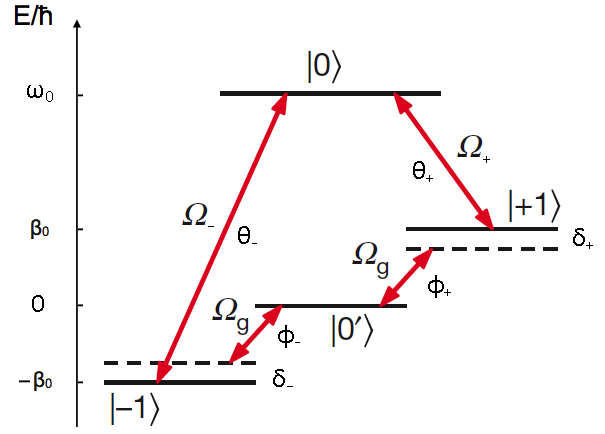}
\caption{Four-level system for the realisation of the dressed state qubit (elements reproduced from \cite{timoney2011quantum}). Couplings in the microwave and radio wave domain are shown ($\Omega_{+/-}$ and $\Omega_g$ respectively). Rabi frequencies are denoted by $\Omega_i$, detunings by $\delta_i$ and phases of the long-wavelength fields by $\theta_i, \phi_i$. Another possible coupling not shown is between $\ket{0}$ and $\ket{0'}$, which is described using $\Omega_z, \theta_z$ and $\delta_z$. States $\ket{-1}$ and $\ket{1}$ are the magnetic-sensitive levels, and the presence of static magnetic field is assumed.}
\label{f1}
\end{figure}

Our scheme retains all the key elements of the original proposal by Timoney et al. \cite{timoney2011quantum} (also described in Webster et al. \cite{webster2013simple}), including initialisation, read-out and encoding of the logical qubit with the help of dressed states. The particular candidate for experimental implementation would be trapped ${}^{171}Yb^+$ ions \cite{wund, webster2013simple}, however, the gate derivations are presented for a generic magnetic-sensitive four-level system, depicted in Figure \ref{f1}. This is done in order to maintain continuity with the work by Timoney et al. \cite{timoney2011quantum} and provide further clarification to the mathematics presented therein. States $\ket{-1}$ and $\ket{1}$ are the magnetic-sensitive levels, and the presence of static magnetic field generates their splitting in energy. The $\ket{-1} \leftrightarrow \ket{1}$ transition is considered forbidden in line with the ${}^{171}Yb^+$ case.

The case of ${}^{171}Yb^+$ is discussed in more detail in Section \ref{F18}, where formulae for the magnitude of the Zeeman splittings in the system are reported, along with relative energy level height. Experimentally, ${}^{171}Yb^+$ ions would be initialised into the state corresponding to $\ket{0}$ by optical pumping, after which a microwave $\pi$-pulse would bring the state to $\ket{-1}$. One creates dressed states by means of a partial STIRAP sequence starting at $\ket{-1}$ using the microwave fields $\Omega_{+/-}$, which is halted in the middle, leaving the fields on at constant strength. Choosing appropriate field phases enables one to reach either of the dressed states:
\begin{align}
\notag &\ket{D}=\frac{1}{\sqrt{2}}(\ket{-1}-\ket{1}) \\
&\ket{B}=\frac{1}{\sqrt{2}}(\ket{-1}+\ket{1}).
\label{d1}
\end{align}
Experimental creation of such states has been achieved using ${}^{171}Yb^+$ ions with lifetimes in excess of $500ms$ \cite{timoney2011quantum, webster2013simple}.

Quantum operations are to be carried out using either $\{\ket{D}, \ket{0'}\}$ or $\{\ket{B}, \ket{0'}\}$ as the logical qubit. The four-state system is viewed in either case by considering the remaining pair of orthogonal states: $\{\ket{B}, \ket{0}\}$ and $\{\ket{D}, \ket{0}\}$, respectively. We also define 'up' and 'down' as alternative basis states, which will be important in the discussion:
\begin{align}
\notag&\text{ \,\,For the D-qubit:} \,\,\,\,\,\,\,\,\,\,\,\,\,\,\,\,\,\,\,\,\,\,\,\, \text{ For the B-qubit:} \\
\notag&\ket{u}=\frac{1}{\sqrt{2}}(\ket{B}+\ket{0}) \,\,\,\,\,\,\,\,\,\,\,\,\,\,\,\, \ket{u}=\frac{1}{\sqrt{2}}(\ket{D}+\ket{0})\\
&\ket{d}=\frac{1}{\sqrt{2}}(\ket{B}-\ket{0}) \,\,\,\,\,\,\,\,\,\,\,\,\,\,\,\,\, \ket{d}=\frac{1}{\sqrt{2}}(\ket{D}-\ket{0}).
\label{d2}
\end{align}

\begin{figure}
\includegraphics[width=0.34\textwidth]{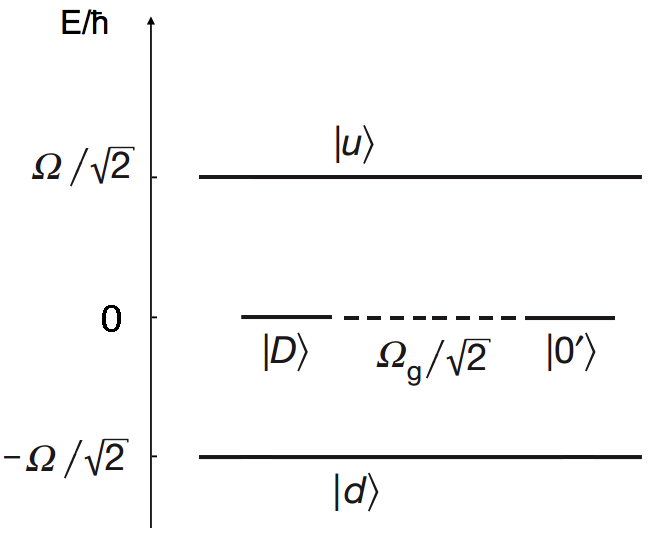}
\caption{Viewing the physical system of Figure \ref{f1} in the dressed state basis (elements reproduced from \cite{timoney2011quantum}). Taking the example of the D-qubit, the arrangement matches equation (\ref{h15}). Microwave dressing fields are held constant at $\Omega_{+/-} = \Omega$, and the dressed states are defined using (\ref{d1}) and (\ref{d2}). Analogous arrangement is found for the B-qubit. The dashed line represents the magnitude of the coupling strength between the qubit states.}
\label{l56}
\end{figure}

During the halted STIRAP sequence, with the dressing fields constant at $\Omega_{+/-} = \Omega$, it is found that $\ket{u}$ and $\ket{d}$ diagonalise the Hamiltonian. Figure \ref{l56} plots the energy level diagram for the D-qubit case, showing how an energy gap is opened between the qubit space and the states $\ket{u}$ and $\ket{d}$, an arrangement which could also be used for qutrit realization \cite{cohen2014proposal}. 

Interactions within the qubit space can be driven by introducing additional radio wave fields (Rabi frequency $\Omega_g$). This arrangement provides the starting point for the single and multi-qubit gates presented in the paper.

It will be illustrated how single and multi-qubit gates can be realised in such a set-up, using, for the multi-qubit case, a magnetic field of constant gradient to strengthen the coupling between neighboring ions. In contrast to recent work, where second-order Zeeman shift is intrinsically used \cite{webster2013simple}, we show how the simple first order shift is sufficient to construct a universal gate set. Further, we ease the experimental requirements by setting equal the phases and detunings of the radio wave fields: $\phi_- = \phi_+$, $\delta_- = \delta_+$. In other words, the radio wave couplings in Figure \ref{f1} would be created by a single field interacting with both $\ket{-1} \leftrightarrow \ket{0'}$ and $\ket{0'} \leftrightarrow \ket{1}$ pairs of levels simultaneously. In the case of the two-qubit gate (Section \ref{b52}), interactions would be created by two radio frequency fields per qubit, which would each interact with both pairs of levels, thus generating four couplings per trapped particle.

Having demonstrated our scheme in detail, we discuss the case of non-linear Zeeman shift, considering modifications of our designs in light of the greater experimental facility  (Section \ref{F18}).

\section{Single-qubit operations}

This Section presents the techniques that enable universal single-qubit rotation to be executed on the dressed state qubit. We propose and describe two distinct gates (Sections \ref{basic}, \ref{adiz}) as well as an adiabatic transfer technique (Section \ref{atran}). Further, we mention two additional single-qubit gate designs, which are described in detail in the appendix. 

Considering the eventual experimental implementation, within the set-up of an ion chain, addressing of individual qubits would be accomplished by separation in frequency space, with the help of static magnetic gradient \cite{mintert2001ion}. This relies on gates coupling only such pairs of levels, where at least one state is magnetically sensitive, so that resonant frequencies vary along the trap axis. The gates proposed in this Section do retain this property.

As the two key limiting factors to gate fidelity, we consider explicitly the noise in the ambient magnetic field and noise due to the instability of the microwave dressing frequencies $\Omega_{+/-}$. It will be shown how these effects can be overcome to reach gate fidelities in excess of $99\%$ in numerical simulation. In order to maintain analytical tractability and illustrate precisely the role of the two sources of experimental noise, the single-qubit gates will be presented and analysed in the slightly simplified set-up with zero magnetic gradient present in the trap. Section \ref{effect} provides justification for regarding the gradient a negligible effect for the single-qubit gates.

\subsection{Hamiltonian and noise sources}

We write down the single-particle Hamiltonian of the most general useful form ($\hbar$ is omitted throughout the paper). We also discuss in this Section the mathematical treatment of the noise sources to be considered explicitly, and briefly describe how the numerical simulations will be run. Figure \ref{f1} defines the phases, detunings and Rabi frequencies used. An extra possibility not drawn for clarity of presentation is the coupling between $\ket{0}$ and $\ket{0'}$, which is defined using $\Omega_z, \theta_z,$ and $\delta_z$. Within the dipole approximation \cite{gerry2005introductory}, one obtains the following expression:
\begin{align}
\label{time-dep}H& =\omega_0\ket{0}\bra{0}+\beta_0\ket{1}\bra{1}-\beta_0\ket{-1}\bra{-1} +  \\
&\,\,\,\,\,\,\,\,\notag \Omega_-\cos{ \bigg((\omega_0 + \beta_0)t+\theta_- \bigg)} \ket{-1}\bra{0} + h.c. +  \\
&\,\,\,\,\,\,\,\,\,\,\,\,\,\,\,\,\,\,\,\,\,\,\,\,\,\,  \Omega_+\cos{ \bigg((\omega_0 - \beta_0)t+\theta_+ \bigg)}\ket{1}\bra{0} + h.c. + \\
&\,\,\,\,\,\,\,\,\notag \Omega_g\cos{ \bigg((\beta_{0}-\delta_-)t+\phi_- \bigg)}\ket{-1}\bra{0'} + h.c. +\\
&\,\,\,\,\,\,\,\,\,\,\,\,\,\,\,\,\,\,\,\,\,\,\,\,\,\,  \Omega_g\cos{ \bigg((\beta_{0}-\delta_+)t+\phi_+ \bigg)}\ket{1}\bra{0'}+h.c. + \\
&\,\,\,\,\,\,\,\, \Omega_z \cos{\bigg((\omega_0 - \delta_z) t+\theta_z \bigg)} \ket{0'}\bra{0} + h.c.
\end{align}
Moving to the interaction picture with respect to the time-independent part (\ref{time-dep}) and performing the rotating wave approximation:
\begin{align}
H= \notag \,\,&\frac{1}{2} \bigg(e^{-i \theta_- }\Omega_-\ket{0}\bra{-1} + e^{-i \theta_+}\Omega_+\ket{0}\bra{1} \,\,+ \\ 
\notag &e^{i \phi_- - i \delta_- t }\Omega_g\ket{-1}\bra{0'}  + e^{i \phi_+ - i \delta_+ t}\Omega_g\ket{0'}\bra{1}  \,\,+ \\ 
\label{ham} &e^{-i \theta_z+i \delta_z t}\Omega_z\ket{0}\bra{0'}  \bigg) + h.c.
\end{align}

Magnetic fluctuations are introduced within the mathematical treatment by considering the following additional term to the Hamiltonian, affecting the two magnetic-sensitive states:
\begin{align}
& \label{noise}\,\,\mu(t) \bigg( \ket{1}\bra{1} - \ket{-1}\bra{-1} \bigg).
\end{align}
Here, $\mu(t)$ is a stochastic process of amplitude proportional to fluctuations in the ambient magnetic field.

Regarding the mathematical treatment of noise in $\Omega_{+/-}$, we approximate and define:
\begin{align}
\notag \Omega_- + \Omega_+ \approx 2\Omega \\
\Omega_- - \Omega_+ = \delta_\Omega
\label{omega}
\end{align}
where $\Omega$ is taken as constant and $\delta_\Omega$ is a second stochastic process. Since the radio frequency couplings will be generated, in the one-qubit case, by a single field only, no analogous term is introduced for $\Omega_g$. In effect, the absence of explicit mathematical treatment of noise in $\Omega_g$ is based on the following approximation:
\begin{align}
\Omega_g + \delta \Omega_g \approx \Omega_g
\end{align}
where $\delta \Omega_g$ is the noise contribution to the single radio frequency field.

The magnitude of the magnetic noise term $\mu$ can be quantified by its standard deviation $SD_\mu$. Section \ref{num}, which discusses the single-qubit numerical simulations in detail, provides an estimate for this parameter, based on experimental measurements, of $2 \pi \cdot 100 \,$ Hz. In numerical simulations, $SD_\mu$ will be set to a constant value, and a particular spectral density profile will be assumed (see Table \ref{tt} and Figure \ref{spec}).

In contrast, the magnitude of $\delta_\Omega$ will be modeled as being proportional to $\Omega$. One can assume normally distributed noise in the strength of the microwave fields $\Omega_{+/-}$ with standard deviation $f \Omega$. Experimentally, $\Omega_{-}$ and $\Omega_{+}$ can be generated from the same microwave source that is multiplied by a radio frequency driving field. In that case, the noise in the microwave Rabi frequencies would be strongly correlated. However, under the extreme assumption of complete independence between $\Omega_{-}$ and $\Omega_{+}$, the standard deviation of $\delta_\Omega$ would equal $\sqrt{2} f \Omega$:
\begin{align}
\notag & SD_\Omega = f \Omega \\
& SD_{\delta_\Omega} = \sqrt{2} f \Omega
\label{SD}.
\end{align}
In the experimental context, correlation between $\Omega_{-}$ and $\Omega_{+}$ would almost certainly reduce the value of $SD_{\delta_\Omega}$ significantly. However, (\ref{SD}) will be used in calculations and numerical simulation because of computational simplicity and for reasons of conservative estimation.

Likewise, the numerical simulations will be run with noise in $\Omega_g$ added, using the same $f$ parameter to quantify its standard deviation:
\begin{align}
SD_{\Omega_g} = f \Omega_g.
\label{SDG}
\end{align}
This is done for reasons of conservative estimation, and also because some higher-order noise effects in the system may manifest as effective noise in the radio frequency field. This aspect will be discussed further in Section \ref{its}.

We acknowledge the existence of other potential sources of experimental noise: phase control, polarisation, possible mismatch between the amplitudes of the two microwave dressing fields. These effects will be discussed furhter in Section \ref{its}, arguing why they are expected to be insignificant.

\subsection{Basic $\sigma_x$/$\sigma_y$ gates}
\label{basic}

Building on the work of Timoney et al. \cite{timoney2011quantum}, it is shown how the $\sigma_y$ gate for the D-qubit and the $\sigma_x$ gate for the B-qubit can be realised by appropriate choice of field phases. Removing the $\ket{0} \leftrightarrow \ket{0'}$ coupling in (\ref{ham}) and choosing:
\begin{align}
\notag & \Omega_{+/-} = \Omega \\
\notag &\theta_+=0  \,\,,\,\,\theta_-=0\\
\notag &\phi_+=\,\,\phi_-=\pi/2\\
&\delta_+=\,\,\delta_-=0
\label{miegs}
\end{align}
one finds
\begin{align}
H=\notag&\,\frac{\Omega}{\sqrt{2}}\, \bigg(\ket{u}\bra{u} - \ket{d}\bra{d}\bigg)\,+ \\
\label{h15} \,&\,\frac{\Omega_g}{\sqrt{2}}\, \bigg(i\ket{D}\bra{0'} - i\ket{0'}\bra{D}\bigg).
\end{align}
And setting:
\begin{align}
\notag & \Omega_{+/-} = \Omega \\
\notag &\theta_+=\pi \,\,,\,\,\theta_-=0\\
\notag &\phi_+=\,\,\phi_-=0\\
&\delta_+=\,\,\delta_-=0
\label{miegs2}
\end{align}
one obtains
\begin{align}
H=\notag&\,\frac{\Omega}{\sqrt{2}}\, \bigg(\ket{u}\bra{u} - \ket{d}\bra{d}\bigg)\,+ \\
\label{h2}\,&\,\frac{\Omega_g }{\sqrt{2}}\, \bigg(\ket{B}\bra{0'} + \ket{0'}\bra{B}\bigg)
\end{align}
using the appropriate definitions of $\ket{u}$ and $\ket{d}$ (\ref{d2}).

It is seen that the radio wave part (Rabi frequency $\Omega_g$) in the above expressions yields the sought-after forms for the single-qubit quantum gates, while microwave dressing fields (Rabi frequency $\Omega$) separate the energies of the remaining pair of basis states. The case of the D-qubit (\ref{h15}) has been plotted in Figure \ref{l56}. The energy gap opened by the microwave fields plays a crucial role in shielding the qubit, particularly against the magnetic noise effects. Such a mechanism is common to all the gates presented in this paper.

Further examination reveals that the requirement to set equal the radio wave phases ($\phi_- = \phi_+$) allows for no other $\sigma_i$ gate to be created using this route for either the B or the D-qubits. The scheme could be generalised to consider superpositions of states $\ket{B}$ and $\ket{D}$, so that the logical qubit would now be represented by $\{\ket{0'}, \cos{\gamma} \ket{B}+\sin{\gamma} \ket{D}\}$. In such a case, a single $\sigma_\gamma$ gate in the $xy$ plane of the Bloch sphere becomes feasible for each choice of $\gamma$. However, the technique allows for no second independent rotation to be achieved for the same definition of the logical qubit. Hence, complementary techniques will be required to realise universal single-qubit rotation.

Considering the D-qubit case and adding the two noise sources (\ref{noise}, \ref{omega}), expression (\ref{h15}) remains unaltered, but it needs to be complemented by the following term:
\begin{align}
\notag H_{n}=\,\,\,&\bigg(-\frac{\mu}{\sqrt{2}}+\frac{\delta_\Omega}{4}\bigg) \ket{D}\bra{u} \,+\, h.c. \,+\\
&\,\,\,\,\,\,\,\,\,\,\,\,\,\,\, \bigg(-\frac{\mu}{\sqrt{2}}-\frac{\delta_\Omega}{4}\bigg) \ket{D}\bra{d}\,+\,h.c.
\label{sunday}
\end{align}
Moving to the interaction picture with respect to the microwave and radio wave part (\ref{h15}), one finds that rotating phases of frequency $(\Omega\pm\Omega_g)/\sqrt{2}$ are introduced to all terms in $H_{n}$ (\ref{sunday}). Therefore, provided that the magnitudes of $\mu, \delta_\Omega$ are much smaller than the rotation frequency, the terms can be deemed negligible within the rotating wave approximation. 

The magnitude of $H_n$ (in the interaction picture) can be further estimated by adiabatic elimination \cite{cohen2006quantum}, writing the time-propagation operator $U(t)$ in orders of $H_{n}$ and looking for terms that grow linearly with $t$ (secular terms). In the second order, one recovers corrections to the energies of $\ket{u}$ and $\ket{d}$, in addition to terms in the qubit space:
\begin{align}
\label{sr7} \notag H_{n2} = &\frac{\mu \Omega \delta_\Omega}{2 (\Omega^2-\Omega_g^2)} \bigg(\ket{D}\bra{D} + \ket{0'}\bra{0'} \bigg)\\
&\,\,\,\,\,\,\,\,\,+i \frac{(8 \mu^2 +\delta_\Omega^2 ) \Omega_g}{8 \sqrt{2} (-\Omega^2+\Omega_g^2)} \bigg(\ket{D}\bra{0'} - \ket{0'}\bra{D} \bigg).
\end{align}
This amounts to an energy shift and a correction to the $\sigma_y$ gate couplings. In the third order, one finds population leakage terms out of the qubit space of magnitude:
\begin{align}
\label{sr8} \frac{\Omega^2 (\sqrt{8}\mu \pm \delta_\Omega)^3}{32(\Omega^2-\Omega_g^2)^2} \,\,\,,\,\,\,\frac{\Omega \Omega_g (\sqrt{8}\mu \pm \delta_\Omega)^3}{32 (\Omega^2-\Omega_g^2)^2}.
\end{align}

Minimisation of these unwanted terms can be accomplished by suppression through large denominator. The conditions for this can be summarised as:
\begin{align}
&\sqrt{|\Omega^2 - \Omega_g^2|} \,\, \gg \,\, \{\mu, \delta_\Omega \}.
\label{constr}
\end{align}

\subsection{Further sources of noise}
\label{its}
In this Section, we examine the basic gate arrangement in more detail, considering the effects of further sources of experimental noise, arguing why they can be treated as negligible. We begin with the issue of phase control.

Noise in the phase of the radio frequency driving field relative to the microwave fields can be described as $\phi_+=\phi_-=\pi/2 +\delta\phi$, with $\delta\phi \ll 1$. Thus, the gate operator (\ref{h15}) becomes:
\begin{align}
H=\notag&\,\frac{\Omega}{\sqrt{2}}\, \bigg(\ket{u}\bra{u} - \ket{d}\bra{d}\bigg)\,+ \\
\label{cls} \,&\,\frac{\Omega_g}{\sqrt{2}}\, \bigg(i (\ket{D} \cos{\delta\phi} + \ket{B} \sin{\delta\phi})\bra{0'} + h.c.\bigg)
\end{align}
In the interaction picture with respect to the dressed state energy, the terms coupling $\ket{B}$ and $\ket{0'}$ rotate fast, since $\Omega_g\delta\phi \ll \Omega$, and can be neglected within the rotating wave approximation. Their contribution results in the following Stark shift:
\begin{equation}
\frac{(\Omega_g\delta\phi)^2}{\sqrt{2}\Omega}\left( \left\vert u\right\rangle \left \langle u \right\vert - \left\vert d\right\rangle \left \langle d \right\vert  \right)
\end{equation}
which adds to the noise in the microwave Rabi frequency. Yet, it is second order in the small parameter $\delta \phi$ and thus can be regarded as negligible in the derivation of the previous Section.

It can be seen that (\ref{cls}) contains another noise effect. Instead of $(i\left\vert D\right\rangle\left \langle 0' \right\vert +h.c)$ we obtain $(i\left\vert D\right\rangle\left \langle 0' \right\vert \cos\delta\phi +h.c.)$. When $\delta\phi \ll 1$ this has only significance in the second order of the small parameter and thus can be neglected. One can also view these results as showing the upper limit to which the system remains protected with respect to a sustained drift in the phase error of the radio frequency field.

We now consider a deviation in the relative phase between the two microwave driving fields: $\theta_+ -\theta_- = \delta\theta$ (considering the D-qubit case). It can be seen in the previous Section (\ref{miegs}, \ref{miegs2}), that the relative phase between the two microwave driving fields determines the gate operator and the basis of the qubit states (B or D-qubit). Calculation reveals that in the new basis caused by the microwave phase mismatch the noise in $\delta\theta$ is translated into noise in $\delta\phi = \delta \theta/2$ and can thus be neglected, based on the arguments already presented. The new basis and thus the new gate operator are now $\delta\theta/2-$rotated with respect to the D-qubit basis and the gate operator, decreasing the process fidelity to $1-(\delta\theta)^2/8$. In our derivation we can neglect this, provided the same microwave driving fields are also used for readout and assuming that $\delta\theta$ has a long correlation time and thus is not changed during the whole experiment. Moreover, the effect can again be neglected simply on the grounds of being second-order in the small parameter.

Secondly, one can consider the issue of microwave polarisation. The microwave driving fields' polarisation mismatch has a similar error contribution to that of a phase deviation. The fields addressing the $\ket{-1} \leftrightarrow \ket{0}$ and $\ket{1} \leftrightarrow \ket{0}$ pairs of levels are ideally linearly polarised along exactly the same axis. A small error in the polarisation alignment of the two driving fields is mapped onto an error in the form of a relative phase $\delta \theta$. These effects have already been discussed.

Thirdly, we consider the effect of a mismatch between the average Rabi frequencies of the two microwave fields. If there is a small imbalance of form $\Delta\Omega=\Omega_+-\Omega_-$, with $\Delta\Omega \ll \Omega$, then (\ref{h15}) yields an additional term:
\begin{equation}
\frac{\Delta\Omega}{2}\left( \left\vert 1\right\rangle \left \langle 0 \right\vert + h.c  \right).
\end{equation}
In the interaction picture with respect to the dressed state energy, this term becomes:
\begin{equation}
\frac{\Delta\Omega}{2\sqrt{2}}\left( \left\vert u\right\rangle \left \langle u \right\vert - \left\vert d\right\rangle \left \langle d \right\vert  \right),
\end{equation}
applying the rotating wave approximation to all the other fast rotating terms. This can be added to the noise in the microwave Rabi frequency, which was discussed above. Together with the ambient magnetic field fluctuations $\mu(t)$, there is another noisy term that survives the rotating wave approximation: $(-\mu(t)\Delta \Omega/2\Omega)\left\vert D\right\rangle \left \langle D \right\vert$. In our derivation we assume that this term is negligible.

In summary, we argue that we have taken into account the only major noise factors in the preceding analysis of the basic single-qubit gate, with other effects being negligible in comparison. Similar arguments can be advanced in the case of the other single-qubit operations.

\subsection{Adiabatic transfer between $\ket{B}$ and $\ket{D}$}
\label{atran}

The basic $\sigma_x$ and $\sigma_y$ gates can be linked for computational purposes by means of population transfer between $\ket{B}$ and $\ket{D}$. This is achieved by adiabatic variation of the microwave phase in a set-up that leaves $\ket{0'}$ decoupled.

Removing the $\ket{0} \leftrightarrow \ket{0'}$ coupling and the radio frequency fields in (\ref{ham}), one sets $\Omega_{+/-}= \Omega$. This provides the timescale on which adiabacity would be maintained. One also sets to zero one of the microwave phases: $\theta_-=0$. The transfer is based on slow variation of the other microwave phase $\theta_+(t)$, such that the system is kept in the zero-eigenvalue state: 
\begin{align}
\ket{\Psi_0(t)}=\frac{1}{\sqrt{2}}\,\,(\ket{-1}-e^{i\theta_+}\ket{1}).
\end{align}
Moving from $\ket{D}$ to $\ket{B}$ is achieved by varying $\theta_+$ from $0$ to $\pi$ and moving from $\ket{B}$ to $\ket{D}$ is obtained by varying the opposite way. Given that $\ket{0'}$ remains decoupled throughout, the following evolutions are enabled:
\begin{align}
\notag a\ket{D}+b\ket{0'} &\longrightarrow a e^{-i \pi/2} \ket{B}+b\ket{0'}\\
a\ket{B}+b\ket{0'} &\longrightarrow a e^{i \pi/2} \ket{D}+b\ket{0'}.
\label{arrow}
\end{align}
The Berry's phase has been added in the expressions above, which can be calculated using standard formulae \cite{berry1, berry2}. In the numerical simulations (Section \ref{num}), we vary the microwave phase continuously over a greater range, which yields an outcome state that is a straightforward linear extension of (\ref{arrow}).

To analyse the effects of noise, one views the system in the adiabatic basis \{$\ket{0'}, \ket{\Psi_0}, \ket{u_{ad}}, \ket{d_{ad}}$\}, where the noiseless Hamiltonian is diagonalised. The states \{$\ket{0'}, \ket{\Psi_0}$\}, which represent the qubit space, lie at zero energy, while the latter two time-dependent orthogonal eigenstates are found to lie at energies $\pm \Omega/\sqrt{2}$. This way, an energy gap is realised.

Applying the appropriate basis change to magnetic noise (\ref{noise}), and introducing effects due to microwave instability (\ref{omega}), one finds the following noise contribution:
\begin{align}
\notag H_n=\,\,\,&e^{-i \theta_+}\bigg(-\frac{\mu}{\sqrt{2}}+\frac{\delta_\Omega}{4}\bigg) \ket{\Psi_0}\bra{u_{ad}} +\,h.c. \, +\\
&\,\,\,\,\,\,\,\,\,\,\,\,\,\,\, e^{-i \theta_+}\bigg(-\frac{\mu}{\sqrt{2}}-\frac{\delta_\Omega}{4}\bigg) \ket{\Psi_0}\bra{d_{ad}}\,+\,h.c.
\end{align}
Moving to the interaction picture with respect to the noiseless Hamiltonian $(\Omega/\sqrt{2}) \cdot(\ket{u_{ad}}\bra{u_{ad}}-\ket{d_{ad}}\bra{d_{ad}})$ will introduce rotations to all terms in $H_n$, making them negligible within the rotating wave approximation for sufficiently large $\Omega$.

Expanding the time-propagation operator in orders of $H_n$ (in the interaction picture) and looking for secular terms, one finds in the second order a term affecting the qubit space:
\begin{align}
H_{n2}= \frac{\mu \delta_\Omega}{\Omega}\ket{\Psi_0}\bra{\Psi_0}.
\end{align}
The third order is found to contain leakage terms out of the qubit space of functional forms: $\mu^3/\Omega^2$, $\mu^2\delta_\Omega/\Omega^2$, $\mu \delta_\Omega^2/\Omega^2$, $\delta_\Omega^3/\Omega^2$. Minimising these unwanted couplings requires:
\begin{align}
\notag &\Omega \gg SD_\mu \\
&f \ll 1.
\label{constr2}
\end{align}

In contrast to the basic $\sigma_i$ gates, where the speed is governed by the radio frequency field strength and the noise suppression criteria only, the maximum speed of adiabatic transfer is governed by $\Omega$, the noise suppression criteria, and the requirement for the evolution to remain adiabatic. The effect of adiabacity constraints will be further illustrated in Section \ref{num}.

\subsection{Adiabatic $\sigma_z$ gate}
\label{adiz}
 
We construct a $\sigma_z$ gate based on adiabatic evolution and the Berry's phase. The gate idea follows the proposal by Duan et al. \cite{duan2001geometric}, although it is modified in important ways to suit the present set-up and improve speed and resilience. The gate is composed of three adiabatic segments consisting, respectively, of: altering the phase of the dressing field, ramping down the dressing field while ramping up the gate field, ramping the gate field down and the dressing field up with a different phase. Detailed explanation in more abstract mathematical terms is supplied in the remainder of this Section.

The gate is illustrated for the case of the D-qubit, noting that analogous construction also exists for the B-qubit. One removes the $\ket{0} \leftrightarrow \ket{0'}$ coupling in (\ref{ham}) and introduces adiabatic variables $R_1(t)$ and $R_2(t)$ as follows: 
\begin{align}
&\notag \Omega_{+/-} = \Omega \sin{(R_2)}\\
&\notag \theta_-=\,\,\theta_+ = R_1 \\
&\notag\Omega_g = \Omega \cos{(R_2)}\\
&\notag\phi_- = \,\phi_+ = 0\\
\label{c1} &\delta_+=\,\,\delta_-=0.
\end{align}
Again, $\Omega$ fixes the adiabatic timescale for the gate. 

Substituting into the noiseless Hamiltonian (\ref{ham}) one obtains the expression:
\begin{align}
H =\,\,\,\, &\frac{\Omega e^{i R_1} \sin{R_2}}{\sqrt{2}} \ket{B}\bra{0} + \frac{\Omega \cos{R_2}}{\sqrt{2}} \ket{B}\bra{0'} \,\,\,+ \,\,\,h.c.
\label{hhh}
\end{align}
It is seen that $\ket{D}$ remains decoupled. The $\sigma_z$ gate will be created by inducing the Berry's phase in the $\ket{0'}$ component, effecting the following evolution:
\begin{align}
a\ket{D}+b\ket{0'} &\longrightarrow a \ket{D}+b e^{i \Phi} \ket{0'}.
\label{arrow2}
\end{align}
This will be enabled by the zero-energy eigenstate of (\ref{hhh}):
\begin{align}
\ket{\Psi_0(t)}=\,\,-e^{-i R_1} \cos{R_2} \,\ket{0} + \sin{R_2}\, \ket{0'}.
\label{circ}
\end{align}

To begin and end at state $\ket{0'}$, any adiabatic evolution of $\ket{\Psi_0(t)}$ in the $\{R_1, R_2\}$ plane will need to begin and end on the line $R_2 = \pi/2$. Moreover, in order to maintain continuity with the basic gate arrangement for the D-qubit case (\ref{miegs}), any viable path will be constrained to begin and end at $A$, in order to have the correct microwave phases. The Berry's phase generated by any such trajectory can be calculated \cite{berry1, berry2}:
\begin{align}
\notag &\Phi=i \int_{{\bf R}_{i}}^{{\bf R}_{f}}  \bigg( \bra{\Psi_0} \partial_{R_1} \ket{\Psi_0} d{R_1} + \bra{\Psi_0} \partial_{R_2} \ket{\Psi_0} d{R_2} \bigg) =\\
&\,\,\,\,\,\,\,\,\,\,\,\,\,\,\,\,\,\,\,\,\,\,\,\,\,\,\,\,\,\, = \int_{{\bf R}_{i}}^{{\bf R}_{f}} (\cos{R_2})^2 d{R_1}.
\label{iint}
\end{align}

\begin{figure}
\includegraphics[width=0.30\textwidth]{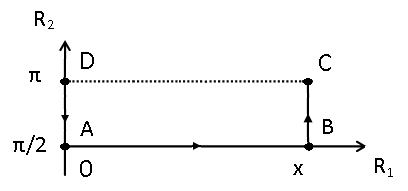}
\caption{Proposed paths of the variables $R_{1,2}(t)$ (\ref{c1}) for the adiabatic $\sigma_z$ gate. The path begins and ends at point $A$, following the arrows. Detailed examination reveals that the segment $C \rightarrow D$ can be omitted, still preserving the following of the adiabatic state.}
\label{fk}
\end{figure}

For the purpose of gate speed, it is desirable to find a path that yields the maximum phase while traversing the least distance. It is seen from (\ref{iint}) that moving along $R_2 = \pi/2$ will generate no phase, although the segment $A \rightarrow B$ is found to be necessary. Figure \ref{fk} shows the path we propose, beginning at point $A$ and ending there as well. Furthermore, the segment $C \rightarrow D$ is omitted, based on mathematical arguments to follow.

One uses (\ref{iint}) to establish that no Berry's phase is generated along the segments $B \rightarrow C$ and $D \rightarrow A$. In contrast, the phase generated along $C \rightarrow D$ is found to be $\Phi = R_1(t) -x$. This cancels exactly the time evolution of $\ket{\Psi_0}$ (\ref{circ}), so that along $C \rightarrow D$ the state follows as:
\begin{align}
\ket{\Psi_0(t)}_{BC}=\,\,e^{-i x} \,\ket{0}
\label{BC}
\end{align}
displaying no time evolution. It is also seen that the Hamiltonian (\ref{hhh}) effects no time evolution for $\ket{\Psi_0}$ along $C \rightarrow D$, irrespective of the range $x$.

These arguments allow one to cut out the segment $C \rightarrow D$ altogether, meaning that a trajectory of shorter length and consequently duration can be traversed to effect the gate. The total phase induced at the end of the path into the $\ket{0'}$ state (see (\ref{arrow2})) is found to be $\Phi = -x$.

For the purpose of noise analysis, the Hamiltonian is diagonalised using the adiabatic basis $\{\ket{D}, \ket{\Psi_0}, \ket{u_{ad}}, \ket{d_{ad}}\}$, where the latter two states are found to lie at energies $\pm \Omega/\sqrt{2}$. Applying the basis change to the noise contributions, the following is found:
\begin{align}
\label{36}& H_n=\,\,\, -\frac{\delta_\Omega \sin{2R_2}}{4 \sqrt{2}} \ket{D}\bra{\Psi_0}\,\,\,+ \,\,h.c.\,\,\, +\\
\notag  &\,\,\,\,\,\,\,\,\,\,\,\,\,\bigg(-\frac{\mu}{\sqrt{2}}+\frac{\delta_\Omega\sin{R_2^2}}{4}\bigg) \ket{D}\bra{u_{ad}}\,+\, h.c. \,+\\
\label{ss} & \,\,\,\,\,\,\,\,\,\,\,\,\,\,\,\,\,\,\,\,\,\,\,\,\,\,\,\,\,\,\,\,\,\,\bigg(\frac{\mu}{\sqrt{2}}+\frac{\delta_\Omega\sin{R_2^2}}{4}\bigg) \ket{D}\bra{d_{ad}}\,+\,h.c.
\end{align}
Line (\ref{36}) yields a first-order noise term within the qubit space that is not correctable by the dressing field. 

After transforming $H_n$ to the interaction picture with respect to the noiseless Hamiltonian $(\Omega/\sqrt{2}) \cdot(\ket{u_{ad}}\bra{u_{ad}}-\ket{d_{ad}}\bra{d_{ad}})$, the following extra contribution is found in the qubit space to second order:
\begin{align}
H_{n2} = \frac{\mu \delta_\Omega \sin{R_2^2}}{\Omega}\ket{D}\bra{D}.
\label{two}
\end{align}
Moreover, leakage terms of forms $\delta_\Omega \mu/\Omega, \delta_\Omega^2/\Omega$ are also recovered.

The dominant noise term is by far (\ref{36}), which can be minimised by requiring good microwave stability ($f \ll 1$), and by lowering $\Omega$ (and hence $SD_{\delta_\Omega}$ (\ref{SD})). Considering the first and second order terms only would suggest that a choice of $\Omega$ as low as possible would minimise these lowest-order noise effects. 

However, the third order analysis reveals terms that grow with reduced $\Omega$. The following is found in the qubit space:
\begin{align}
H_{n3}=\,\,\, \frac{\delta_\Omega (8 \mu^2 + \delta_\Omega^2 \sin{R_2}^4)\sin{2 R_2}}{16\sqrt{2}\,\Omega^2} \,\ket{D}\bra{\Psi_0}\,\,\,+ \,\,h.c.
\end{align}
 In addition, leakage terms of the following form are found: $\mu^3/\Omega^2, \mu^2 \delta_\Omega/\Omega^2, \mu \delta_\Omega^2/\Omega^2, \delta_\Omega^3 /\Omega^2$. The requirement to maintain negligible terms such as $\mu^3/\Omega^2$ sets a lower limit on $\Omega$, suggesting the existence of an optimal microwave dressing frequency. Noise minimisation would therefore be achieved, based on these mathematical arguments alone, for:
 \begin{align}
\notag &\Omega = \Omega_{opt} \\
&f \ll 1.
\end{align}
It will be shown in Section \ref{numz} how a value for $\Omega_{opt}$ does indeed emerge numerically for some sets of simulation parameters. A further lower limit on $\Omega$ would be set by the desired gate speed and the adiabacity requirement. It will be illustrated by the numerical simulation how the adiabacity requirement combines with noise effects to determine the attainable fidelity of the gate.

\subsection{Other $\sigma_z$ gate designs}

For completeness, other ways to realise the $\sigma_z$ gate are briefly described, taking the example of the D-qubit. Firstly, it is possible to construct the adiabatic $\sigma_z$ gate via two alternative routes. Section \ref{adiz} has demonstrated how a phase in $\ket{0'}$ can be induced by employing couplings of the following form: $\ket{0'}\leftrightarrow \ket{B}\leftrightarrow \ket{0}$ (see (\ref{hhh})). Alternatively, one can induce the Berry's phase in $\ket{0'}$ by employing couplings of form $\ket{0'}\leftrightarrow \ket{0}\leftrightarrow \ket{B}$, in a set-up that uses $\Omega_{+/-}$ and $\Omega_z$ microwave fields. It is also possible to follow more closely the original proposal of Duan et al. \cite{duan2001geometric}, using the following couplings: $\ket{-1}\leftrightarrow \ket{0}\leftrightarrow \ket{1}$. This arrangement would require microwave fields $\Omega_{+/-}$ only and work by inducing a phase in $\ket{D}$. The disadvantages found for these alternative schemes include lower gate speed, less favourable noise effects, and the need to couple two magnetically insensitive levels. The adiabatic gate presented in Section \ref{adiz} is found to possess the most favourable overall qualities. However, it is also acknowledged that other functional forms for introducing the adiabatic variables $\{R_1, R_2\}$ could be explored.

Secondly, it is also possible to use the effect of Stark shift \cite{james2007effective} to create the $\sigma_z$ gate, a viable alternative to the adiabatic approach. We show two such designs in appendices \ref{DSS} and \ref{RSS}, the first of which relies on detuned $\ket{0}\leftrightarrow \ket{0'}$ coupling. It is shown how microwave dressing can be applied in such a case to shield the gate. The scheme would have the potential disadvantages of having to couple two magnetically insensitive levels, as well as having tighter experimental constraints on the parameters (\ref{SSc}). Likewise, we present in  appendix \ref{RSS} a radio wave Stark shift $\sigma_z$ gate that relies on $\{\phi_- \neq \phi_+\,,\, \delta_- \neq \delta_+\}$, which goes beyond the experimental limitations considered. The gate is added in light of extending the discussion to non-linear Zeeman regime (Section \ref{F18}), and is found to possess good shielding properties.

\subsection{Numerical simulation}
\label{num}

\begin{figure}
\includegraphics[width=0.35\textwidth]{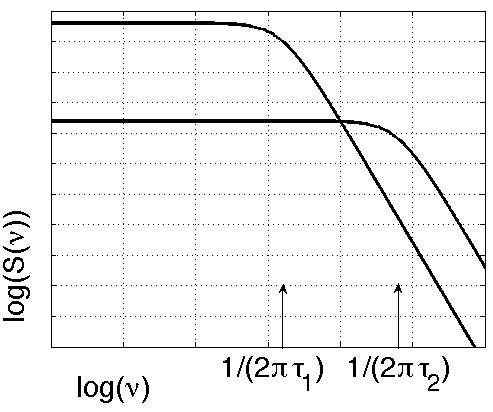}
\caption{Spectral density function for the fully relaxed Ornstein-Uhlenbeck process, plotted in log-log coordinates (formulae taken from \cite{gill2}). Two values for $\tau$ are chosen, illustrating the resulting change in frequency at the turning point. Both plots are normalised to the same total standard deviation.}
\label{spec}
\end{figure}

This Section presents the results of simulating numerically the proposed single-qubit gates, introducing noise in the ambient magnetic field as well as in the Rabi frequencies of the microwave and radio frequency sources. Noise is modeled as the Ornstein-Uhlenbeck (OU) process, using formulae found in Gillespie et al. \cite{PhysRevE.54.2084}. The OU process is stationary, Gaussian and Markovian, and can be thought of as being generated by the interplay between purely random driving and a damping effect. For example, any rectilinear velocity component of a massive Brownian particle (at non-zero temperature and coefficient of diffusion) can be modeled as the OU process \cite{PhysRevE.54.2084}. Figure \ref{spec} plots the spectral density function that the OU process gives rise to.

Two parameters need to be fixed in order to specify fully the time evolution for the OU process: the relaxation time $\tau$ and the diffusion constant $c$. The standard deviation for the fully relaxed OU process is given by $\sqrt{c \tau/2}$. This provides the first useful physical constraint. The second is found by considering the spectral density function of the OU process, plotted in Figure \ref{spec}. The plots are made in log-log coordinates using two choices of $\tau$ and normalising to the same total power. As can be seen, $\tau$ parametrises the turning point for the spectral density function, and this provides the second useful link to physical observables.

As for the Rabi frequency noise, we have already introduced its standard deviation using the fractional $f$ parameter (\ref{SD}, \ref{SDG}). In simulations, a range of $f$ between $0.01$ and $0.05$ will be explored. The relaxation time for the Rabi frequency noise will be labeled as $\tau_f$. We consider a pair of $\tau_f = \{3.2,  32\}$ ms, which corresponds to the turning point frequency lying between $5$ and $50$ Hz, in anticipation that Rabi frequency noise would be dominated by lower frequencies.

For the estimation of magnetic noise parameters, we consider firstly the measurement by Timoney et al. (the preprint version) \cite{T2}, where the lifetime of the dressed state $\ket{D}$ is reported to be $1700 \pm 300$ms in the presence of the microwave dressing fields of strength $\Omega = 2\pi \cdot 36.5$ kHz. We also consider direct measurements of the spectral density function of magnetic noise provided informally by the experimental group of Wunderlich at Siegen \cite{wund}, displaying an overall shape broadly consistent with the OU model (Figure \ref{spec}). These measurements suggest that the relaxation time for magnetic noise (labeled $\tau_\mu$) in the range of $0.1$ms would be a good estimate. We extend the range for our simulations to $\tau_\mu = \{0.016,  0.16\}$ ms, corresponding to the turning point frequency occurring at 1 to 10 kHz.

The lifetime measurement by Timoney et al. can be used to gain an estimate for the standard deviation of magnetic noise (labeled $SD_\mu$). Simulating numerically the lifetime of the $\ket{D}$ state, using $f=0.01, \tau_f = 3.2$ ms, $\tau_\mu = 0.1$ ms leads to results that are consistent with the Timoney measurement for $SD_\mu = 2 \pi \cdot 100$ Hz. Substituting $\tau_f = 32$ ms into the simulation leads to $\approx10\%$ improvement in the lifetime of the $\ket{D}$. We run the simulations using the range $SD_\mu = \{100,  500\}$ Hz to explore a broader range.

Table \ref{tt} provides a summary of the combinations of noise parameters used and their respective colour markers. We have also included a zero-noise entry (black), and quoted the number of runs the simulations will be averaged over.

\begin{figure}
\includegraphics[width=0.46\textwidth]{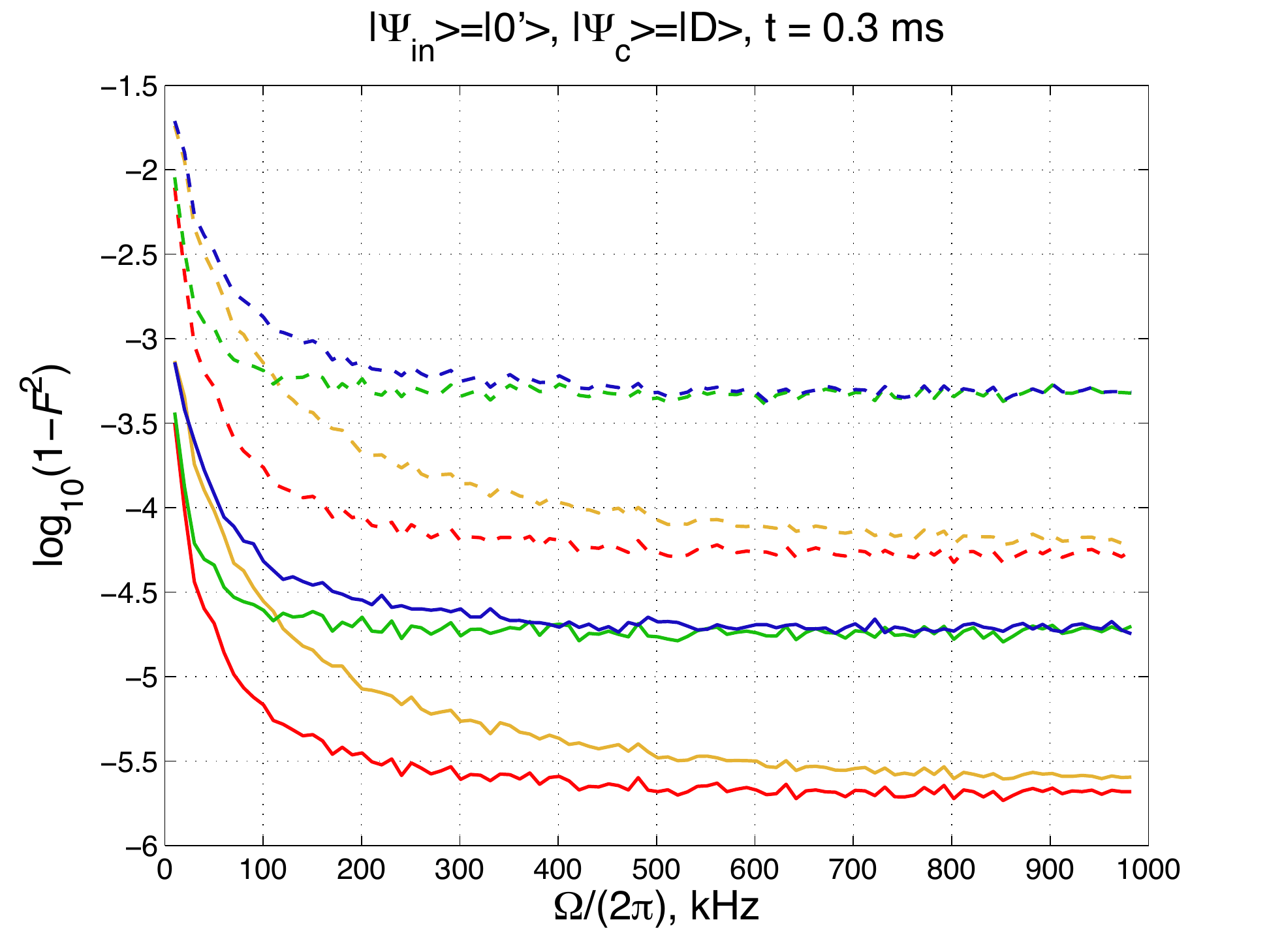}
\includegraphics[width=0.46\textwidth]{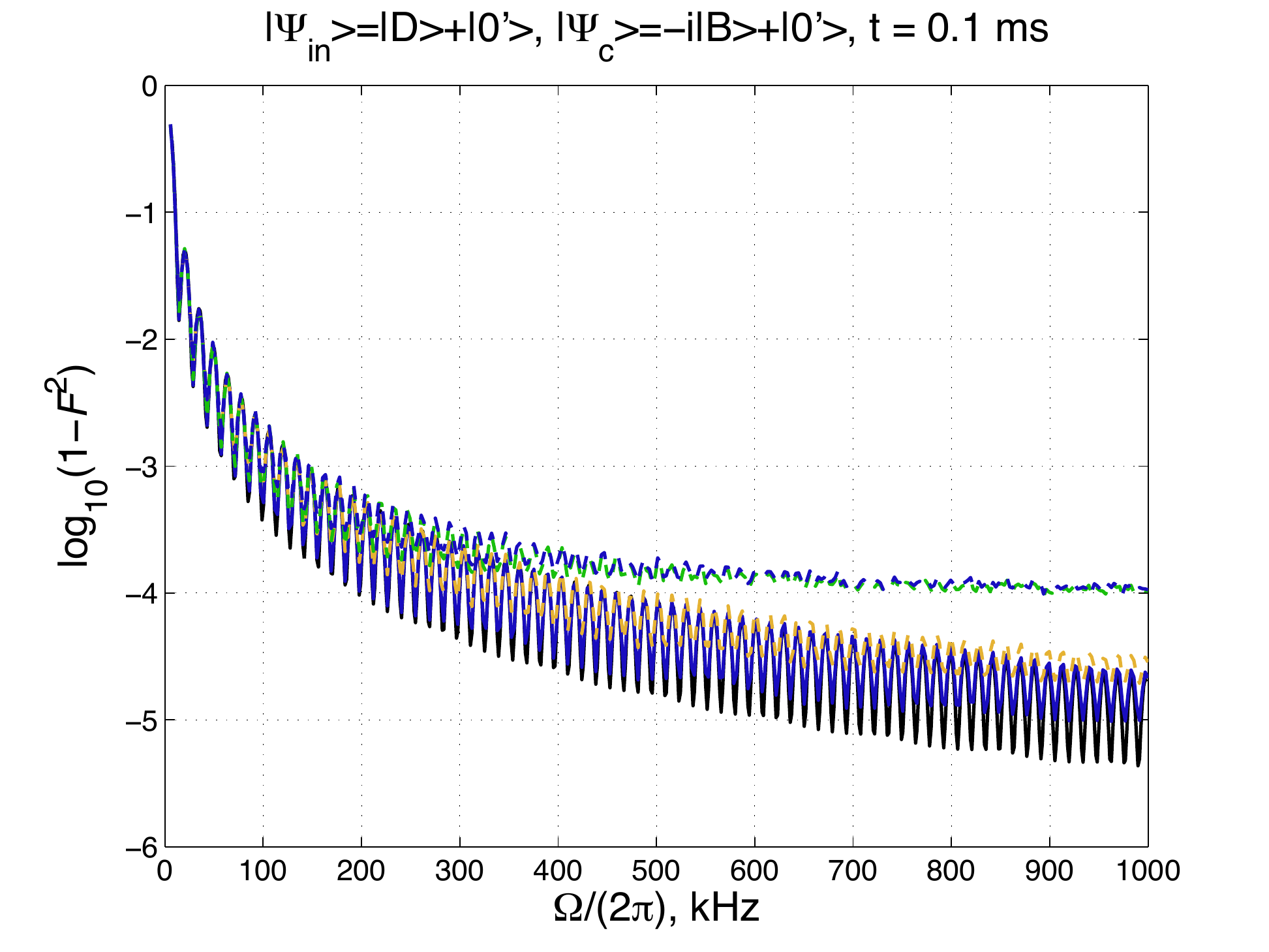}
\includegraphics[width=0.46\textwidth]{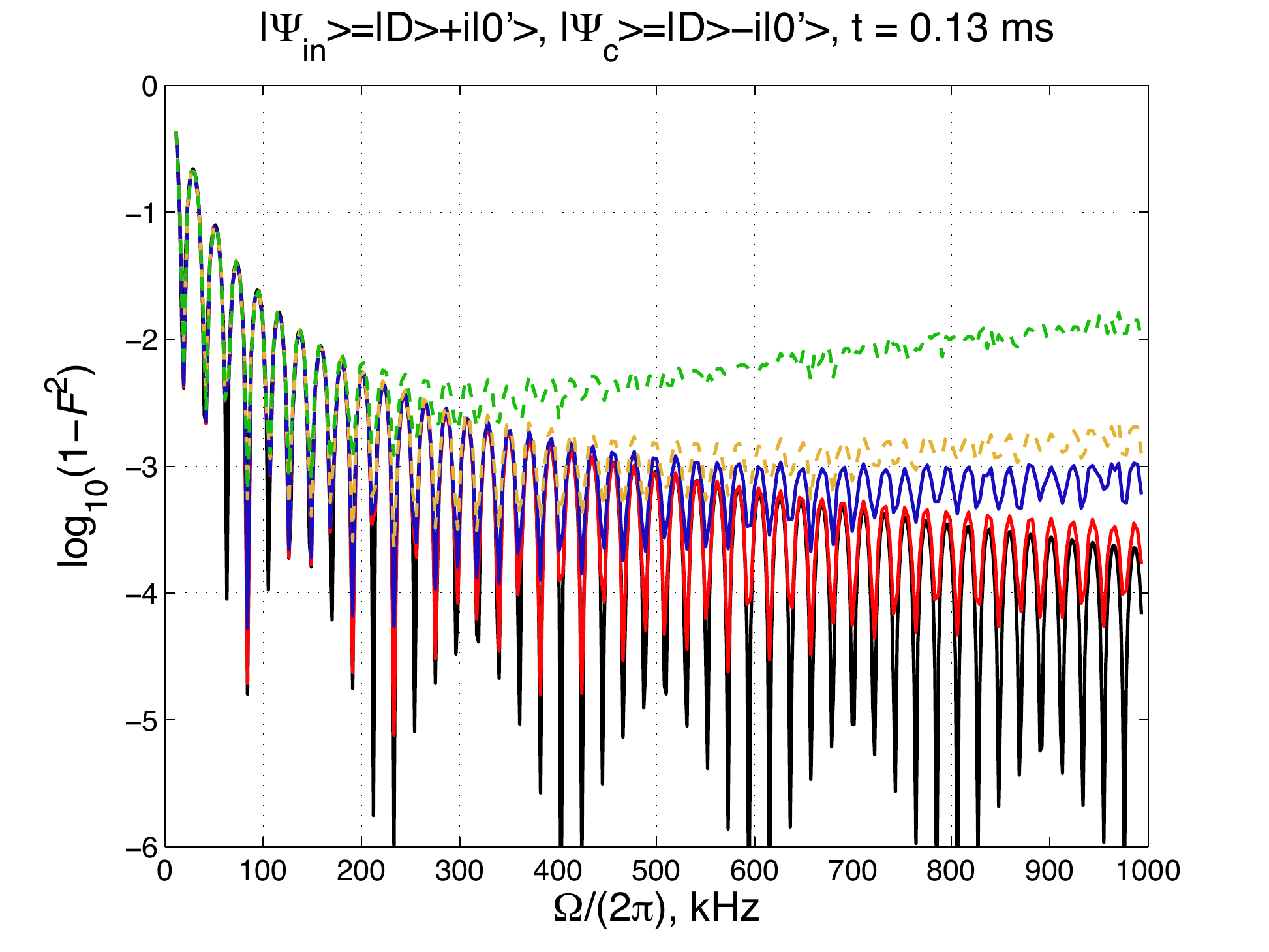}
\caption{Simulation results for the single-qubit operations. $log_{10}(1-F^2)$ is plotted for each process after a $\pi$-pulse (a single Rabi flop), as dependent on the dressing field strength $\Omega$. The (unnormalised) input and comparison states are shown, as are the times for each of the operations. Table \ref{tt} provides the explanation for the colour markers. Simulations with noise present (colour) are displayed after averaging over 200 runs. \underline{TOP}: Basic $\sigma_y$ gate. Other parameters: $\phi_-$= $\phi_+$=1.5708 rad, $\Omega_g$=$2\pi\cdot 1.1785$ kHz. \underline{MIDDLE}: Adiabatic transfer. Adiabatic rate is set to 31.416 rad/ms. \underline{BOTTOM}: Adiabatic $\sigma_z$ gate. Adiabatic rate is set at 47.124 rad/ms. The microwave phase parameter $x=3.1416$ rad.}
\label{f2}
\end{figure}

The fidelity of a quantum state $\rho$, with respect to a desired target or comparison state $\ket{\Psi_c}$, is defined \cite{nielsen2010quantum}:
\begin{align}
F(\ket{\Psi_c}, \rho) = \sqrt{\bra{\Psi_c}\rho\ket{\Psi_c}}
\label{fidl}
\end{align}
so that the probability of finding $\ket{\Psi_c}$ upon measurement is given by $F^2$. There is a square root difference between this definition and the convention used in the paper by M{\o}lmer and S{\o}rensen \cite{sorensen2000entanglement}. As the key figure of merit, we consider the following quantity:
\begin{align}
M = log_{10}(1-F^2)
\label{FM}
\end{align}
which enables the quantification of fine deviations from $F=1$. For example, state fidelity $F=99.99\%$ yields $M=-3.7$ and $F=99.9\%$ yields $M=-2.7$. These figures are close to the most often quoted targets for fault-tolerant quantum computation \cite{steane2003overhead, knill2005quantum}. For each of the single-qubit operations, we evolve the initial state through a single Rabi flop (a $\pi$-pulse) and then find $M$ computationally with respect to the appropriate target state.

\begin{table}
\renewcommand{\arraystretch}{1.5}
\begin{center}
  \begin{tabular}{ c | c | c | c| c | c }
    \hline
     marker & $SD_\mu$, Hz & $\tau_\mu$, ms & $f$ & $\tau_f$, ms & runs averaged  \\ \hline \hline
    black & - & - & - & - & 1 \\ \hline
    red & 100 & 0.16 & 0.01 & 32 & 200 \\ \hline
    yellow & 100 & 0.016 & 0.01 & 32 & 200 \\ \hline
    green & 100 & 0.16 & 0.01 & 3.2 & 200 \\ \hline
    blue & 100 & 0.016 & 0.01 & 3.2 & 200 \\ \hline
    red dashed & 500 & 0.16 & 0.05 & 32 & 200 \\ \hline
    yellow dashed & 500 & 0.016 & 0.05 & 32 & 200 \\ \hline
    green dashed & 500 & 0.16 & 0.05 & 3.2 & 200 \\ \hline
    blue dashed & 500 & 0.016 & 0.05 & 3.2 & 200 \\ \hline
    \hline
  \end{tabular}
\end{center}
\caption{Noise parameters and colour markers for the single-qubit simulations.}
\label{tt}
\end{table}

\subsubsection{Basic $\sigma_x$/$\sigma_y$ gates}
\label{numB}

The value of $M = log_{10}(1-F^2)$ versus dressing field strength for the basic $\sigma_y$ gate is plotted in Figure $\ref{f2}$ top, which should be viewed in conjunction with Table \ref{tt}. It is seen that increased dressing field does indeed provide progressively better shielding up to around $\Omega = 2\pi \cdot 500$ kHz, beyond which a settled value for $M$ is reached. The absence of further improvement can be explained by the existence, within the Hamiltonian noise contribution (\ref{sr7}, \ref{sr8}), of terms that do not diminish with increased $\Omega$ and therefore amount to an uncorrectable effect.

The plot also enables the comparison of the effects of different noise sources. For the majority of the set of parameters considered within the simulation, it is seen that a figure of $M < -3.7$ can be reached. Running the simulation without any noise effects leads to values of M in the vicinity of $-12$, which probably represents the computational limit of the computing package.

The same order of accuracy, as measured by $M$, for the differently coloured plots (red-yellow-green-blue) is found for all three of the single-qubit operations. In the case of the adiabatic operations, less than eight plots will be reproduced for reasons of clarity of presentation.

The time-scale for the basic single-qubit gate (0.3 ms) has been chosen to match that of the two-qubit entangling gate (Section \ref{2qsim}), because aspects of the shielding mechanism are directly analogous in both cases. Further speeding up of the single-qubit gate can be accomplished by increasing $\Omega_g$, while maintaining the constraint for noise suppression (\ref{constr}). The time-scale of $\mu s$ can easily be reached, where the noise effects are even less detrimental.

\subsubsection{Adiabatic transfer}

Figure $\ref{f2}$ middle shows the simulation results for adiabatic transfer using a superposition state. In contrast to the basic gate, a prominent further constraint on gate fidelity is imposed by the adiabatic limit (the black curve), which reduces gate fidelity even in the absence of further noise effects. The total noise contribution is a mixture of effects due to imperfect adiabacity, and the effects of magnetic and Rabi frequency noise. In cases where the latter effects are minimal, the whole noise contribution is dominated by the disturbance due to non-adiabacity, as can be seen in the case of the blue curve in the Figure. The red curve (not plotted) was found to merge even more closely with the adiabatic limit in the simulation.

Performing the operation faster results in the adiabatic limit moving upwards in the plot. In contrast, performing the gate slower would result in less stringent adiabatic limit, but the effects of other factors of noise become more prominent. These considerations, together with any experimental limitation on the maximum $\Omega$ attainable, limit the fidelity of the adiabatic transfer operation. For maximum $\Omega$ of $2 \pi \cdot 1000$kHz, the case plotted (t = $0.1$ ms) represents close to the maximum attainable speed, depending on the real noise conditions.

Conditional on negligible additional magnetic and Rabi frequency noise, the Figure also suggests that a significant gain in $M$ would be possible through precise adjustment of the dressing field Rabi frequency, in order to position oneself in the minimum of the oscillating adiabatic limit. This technique would certainly merit further exploration from an experimental and theoretical point of view. We present the oscillating nature of the adiabatic limit as a computational result and leave its theoretical explanation to further research.

\subsubsection{Adiabatic $\sigma_z$ gate}
\label{numz}

Figure \ref{f2} bottom panel displays results for the adiabatic $\sigma_z$ gate. Again, one sees the combined effects of the magnetic and Rabi frequency noise as well as the adiabatic limit acting as a constraining factor to the gate fidelity. In the case of low magnetic/Rabi frequency noise, the detrimental effects are dominated by the adiabatic limit, as is the case for the red curve. The adiabatic limit is also found to be more complex than in the case of adiabatic transfer, showing further structure and greater amplitude of oscillations. This can be explored further theoretically, noting that the adiabatic path taken to realise the $\sigma_z$ gate is more complex as well. Again, it is of experimental interest to position oneself within a minimum of the adiabatic limit oscillations, thereby effecting a significant improvement in $M$.

For the case where magnetic/Rabi frequency noise amounts to a strong effect (the green curve), evidence can be found for the emergence of optimum dressing frequency $\Omega_{opt}$, thereby confirming the theoretical analysis. 

Again, the case plotted (t = $0.13$ ms) probably represents close to the maximum attainable speed for maximum dressing-field strength of $\Omega = 2 \pi \cdot 1000$kHz.

\subsection{The effect of magnetic gradient}
\label{effect}

The multi-qubit entangling gate presented in Section \ref{b52} makes intrinsic use of static magnetic-field gradient being present along the trap axis. This is also likely to be the case, within the experimental context, for the single-qubit gates. However, introducing a magnetic-field gradient in the single-qubit analysis of the present Section is not expected to add a significant effect.

One can estimate analytically the magnitude of this contribution. Assuming a single motional mode only, the two phonon terms that would be added to the single-qubit Hamiltonian H (\ref{ham}) are:
\begin{align}
H_p = \nu b^\dagger b + \kappa \sigma_z (b^\dagger + b)
\label{pp}
\end{align}
(see (\ref{cc3}) and the definitions (\ref{defs})). No sideband coupling is employed for the single-qubit gates, and one can view the total Hamiltonian in the interaction picture with respect to $\nu b^\dagger b$. This leaves the terms in $H$ unaffected. Evaluating the magnitude of $\kappa \sigma_z (b^\dagger + b)$ after the interaction picture, one recovers the following term in the second order:
\begin{align}
H_{p2} = - \eta^2 \nu \,\,\ket{D}\bra{D}.
\end{align}
This would amount to a tiny effect for realistic experimental parameters (\ref{params3}). The effect on this term of a further interaction picture with respect to the microwave energy gap of form $\Omega \, (\ket{u} \bra{u} - \ket{d} \bra{d} )$ can be neglected, provided that $\Omega \ll \nu$.

Numerical simulation of single-qubit gates with magnetic gradient present has also been carried out to establish that the gradient amounts to a negligible effect.

\section{Multi-qubit gate}
\label{b52}

It is now shown how the dressed-state approach, combined with magnetic-gradient-induced coupling \cite{mintert2001ion}, enables the realisation of an entangling gate. We consider the additional effect of static magnetic-field gradient along the trap axis and show how a Hamiltonian of Jaynes-Cummings form \cite{jaynes1963comparison} can be obtained. It is then used to obtain the fast M{\o}lmer-S{\o}rensen gate \cite{sorensen2000entanglement}.

Magnetic noise effects are discussed explicitly, demonstrating how microwave shielding can be accomplished. We also comment on the detrimental effects due to ion heating and include this process in the numerical simulation. As the third key factor affecting gate fidelity, we consider explicitly the effects of spurious couplings and resonances arising from the system Hamiltonian (\ref{thor}, \ref{cc3}) and the presence of an unused motional mode. Strategies for minimising these unwanted interactions are discussed.

We present the key formulae governing the gate properties and write down explicitly the major parameter constraints arising from the need to minimise spurious coupling effects. Our numerical simulation in Section \ref{2qsim} presents one possible choice of parameters to overcome such effects and yields gate fidelity of up to $F=99.9\%$ (with the noise sources considered). We acknowledge that experimental values can be found to generate even higher fidelities, especially given the number of free parameters to be set (see (\ref{params3})). Further research effort, particularly in light of a potential experimental realisation would be encouraged to find the optimal choice. 

Detrimental effects due to other processes such as noise in the Rabi frequencies, effects due to stray addressing of individual particles in the frequency space, noise in the trap frequency, or known approximations of the trapped-ion physical system \cite{james1998quantum} could also be tackled in future research.

We derive and simulate an entangling gate for the two-particle case with the simplification of considering explicitly a single motional mode only. The issue of avoiding coupling to the other motional mode is discussed, as well as the scope for extending the discussion to the multi-particle case.

\begin{figure}
\includegraphics[width=0.42\textwidth]{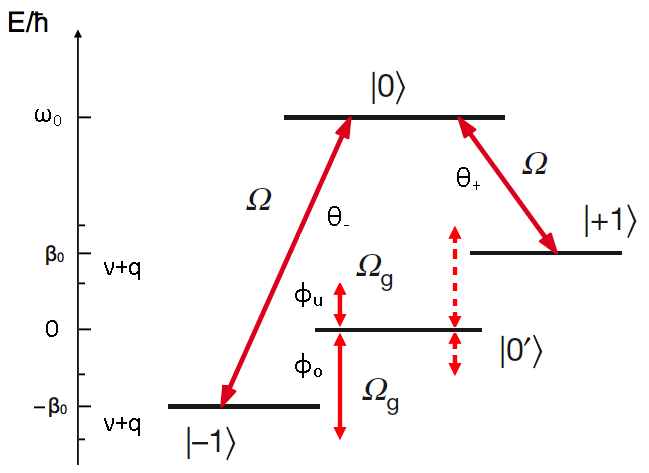}
\caption{Realising the entangling gate (elements reproduced from \cite{timoney2011quantum}). Two radio frequency fields of Rabi frequency $\Omega_g$, detuned by $\pm q$ from the motional sidebands, generate four couplings between the states $\ket{-1}$, $\ket{0'}$, and $\ket{1}$. Microwave fields of Rabi frequency $\Omega$ contribute to the shielding.}
\label{f4}
\end{figure}

\subsection{Set-up and definitions}

Figure \ref{f4} depicts the arrangement for the gate implementation, together with definitions of the microwave and radio frequency fields. Two detuned radio frequency fields are employed, which generate four couplings in the $\{\ket{-1}, \ket{0'}, \ket{+1}\}$ triplet of states. The two microwave fields required will be shown to generate a shielding effect directly analogous to that in the single-qubit gates. The presence of the magnetic gradient makes the energies of $\ket{-1}$ and $\ket{+1}$ position-dependent, so that $\beta_0$ now represents the equilibrium value of $\beta(z)$ for each trapped particle. Communication between individual qubits will be accomplished by means of the shared motional mode of the ions in the trap. The following additional variables are introduced:
\begin{align}
\notag & q - \text{sideband detuning of the RF fields (see Figure \ref{f4})}\\
\notag& \nu - \text{frequency of the shared motional mode}\\
\notag& n - \text{phonon number}\\
\notag& b^\dagger, b - \text{phonon operators, later redefined as:} \\
\notag & \,\,\,\,\,\,\,\,\,\,\,\,\,\,\,\,\,\,\,\,\, \tilde{b}^\dagger= e^{i \nu t}b^\dagger\,\,\,,\,\,\,\tilde{b}= e^{-i \nu t} b\\
\notag& \kappa - \text{constant proportional to the magnetic gradient}, \\ 
\notag& \,\,\,\,\,\,\,\,\,\,\,\,\,\,\,\,\,\,\,\,\,\text{defined explicitly in appendix \ref{etad}}\\
\notag& \eta = \kappa/\nu - \text{the effective Lamb-Dicke parameter} \\
\notag& R - \text{integer parameter characterising the fast} \\
\notag& \,\,\,\,\,\,\,\,\,\,\,\,\,\,\,\,\,\,\,\,\,\text{M{\o}lmer-S{\o}rensen gate (see (\ref{cond1}))}\\
\notag& \sigma_z = \ket{1}\bra{1} - \ket{-1}\bra{-1} \\
\label{defs} & \sigma_+ = \ket{D}\bra{0'} \,\,\,\,,\,\,\,\, \sigma_- = \ket{0'}\bra{D}.
\end{align}

\subsection{Single-particle Hamiltonian}

In the interaction picture with respect to $H_0 = \omega_0 \ket{0}\bra{0} + \beta_0 \ket{1}\bra{1} - \beta_0 \ket{-1}\bra{-1}$ and after performing the rotating wave approximation, one obtains the following Hamiltonian for the interactions depicted in Figure \ref{f4}:
\begin{align}
\notag & H = \,\, \frac{\Omega}{2} \bigg( e^{-i\theta_-} \ket{0}\bra{-1} + e^{-i \theta_+} \ket{0}\bra{1}+h.c. \bigg)\,\,+\\
\notag&\frac{\Omega_g}{2} \bigg(  e^{i (\nu+q) t} e^{-i\phi_u} \ket{0'}\bra{-1} +          e^{- i (\nu+q) t} e^{-i\phi_o} \ket{0'}\bra{-1} + h.c.    \bigg)\,\,+\\
& \label{thor} \frac{\Omega_g}{2} \bigg(    e^{i (\nu+q) t} e^{-i \phi_u} \ket{1}\bra{0'} +  e^{- i (\nu+q) t} e^{-i \phi_o} \ket{1}\bra{0'} + h.c.    \bigg)\\
&\label{cc3}\,\,\,\,\,\,\,\,\,\,\,\,\,\,\,\,\,\,\,\,\,\,\,\,\, +\,\,\nu b^\dagger b + \kappa \sigma_z (b^\dagger + b).
\end{align}
The microwave and radio wave part (\ref{thor}) is directly analogous to the one previously quoted (\ref{ham}). Line (\ref{cc3}) contains the phonon energy and the term due to the presence of the magnetic gradient \cite{mintert2001ion}. 

As the next step, one applies the Schrieffer--Wolff transformation \cite{bravyi2011schrieffer} of form:
\begin{align}
M\rightarrow e^{\eta \sigma_z (b^\dagger - b)}Me^{- \eta \sigma_z (b^\dagger - b)}.
\label{SWT}
\end{align}
Its effect is to introduce factors to all terms in (\ref{thor}) as well as to remove the $\kappa \sigma_z (b^\dagger + b)$ contribution. The following additional term is obtained after the transformation: 
\begin{align}
H_{SW}= - \eta^2 \nu \sigma_z^2. 
\end{align}
The operator $\sigma_z^2$ is not equal to the identity here, since the Hilbert space contains four levels. Moving to the interaction picture with respect to the phonon term $\nu b^\dagger b$, one recovers the following Hamiltonian:
\begin{align}
\notag H & = \,\,\,\frac{\Omega}{2} \bigg( e^{-i\theta_-} \ket{0}\bra{-1}e^{\eta (\tilde{b}^\dagger-\tilde{b})} +h.c. \bigg)\,\,+\\
\notag & \,\,\,\,\,\,\,\,\,\,\,\,\,\,\,\,\,\,\,\,\,\, \frac{\Omega}{2} \bigg( e^{-i \theta_+} \ket{0}\bra{1} e^{-\eta (\tilde{b}^\dagger-\tilde{b})}+h.c. \bigg)\,+\\
\notag&\,\,\,\,\,\,\,\,\,\frac{\Omega_g}{2} \bigg( e^{i (\nu+q) t} e^{-i\phi_u} \ket{0'}\bra{-1} e^{\eta (\tilde{b}^\dagger-\tilde{b})} +h.c. \bigg)\,+\\
\notag&\,\,\,\,\,\,\,\,\,\,\,\,\,\,\,\,\,\,\,\,\,\,\frac{\Omega_g}{2} \bigg( e^{- i (\nu+q) t} e^{-i\phi_o} \ket{0'}\bra{-1} e^{\eta (\tilde{b}^\dagger-\tilde{b})} +h.c. \bigg)\,+\\
\notag&\,\,\,\,\,\,\,\,\, \frac{\Omega_g}{2} \bigg( e^{i (\nu+q) t} e^{-i \phi_u} \ket{1}\bra{0'} e^{\eta (\tilde{b}^\dagger-\tilde{b})} +h.c. \bigg)\,+\\
\notag& \,\,\,\,\,\,\,\,\,\,\,\,\,\,\,\,\,\,\,\,\,\,\frac{\Omega_g}{2} \bigg( e^{- i (\nu+q)  t} e^{-i \phi_o} \ket{1}\bra{0'} e^{\eta (\tilde{b}^\dagger-\tilde{b})} +h.c.  \bigg)\,+\\
\label{sat}&\,\,\,\,\,\,\,\,\,\,\,\,\,\,\,\,\,\,\,\,\,\,\,\,\,\,\,\,\,\,\,\,\,\,\,\,\,\,\,\,\,\,\,\,\,\,\,\,\,\,\,\,\,\,\,\,\,\,\,\,\,\,\,\,\,\,\,\,\,\,\,\,\,\,\,\,\,\,\,\,\,\,\,\,H_{SW}.
\end{align}

\subsection{Jaynes-Cummings form}

The gate will be illustrated for the case of the D-qubit, noting that an analogous construction for the B-qubit is possible. One sets in (\ref{sat}):
\begin{align}
\notag &\theta_- = \theta_+ = 0\\
\label{c2} &\phi_u=\phi_o = 0.
\end{align}
Expanding the coupling terms to first order in $\eta$ and changing basis to $\{\ket{u}, \ket{d}, \ket{D}, \ket{0'}\}$, one obtains:

\begin{align}
\label{line1} H= \,\,&\frac{\Omega}{\sqrt{2}}\,\, \bigg(\ket{u} \bra{u} - \ket{d} \bra{d} \bigg) \,\,+ \\   
\label{qbit} & \frac{\eta \Omega_g}{\sqrt{2}}\,\, \bigg(-e^{i q t}\, b\,\, + \,\, e^{-i q t} \, b^\dagger \,\, \bigg) \, \ket{0'}\bra{D} \,+\, h.c.  \\ 
\label{qbit2} & \frac{\eta \Omega_g}{\sqrt{2}}\,\, \bigg( -e^{-i q t-2i\nu t}\, b\,\, +\,\, e^{i q t+2 i \nu t }\, b^\dagger \,\, \bigg) \, \ket{0'}\bra{D} \,+\, h.c.  \\ 
\label{hres}& \,\,\,\,\,\,\,\,\,\,\,\,\,\,\,\,\,\,\,\,\,\,\,\,\,\,\,\,\,\,\,\,\,\,\,\,\,\,\,\,\,\,\,\,\,\,\,\,\,\,\,\,\,\,\,\,\,\,\, + \,\,H_{res}\,\,+\,\,H_{SW}.
\end{align}

Line (\ref{qbit}) gives the sought-after Jaynes-Cummings type of coupling in the qubit space. The terms oscillating with frequency $\pm q$ will be used in building the entangling gate, while the effect of the faster-oscillating $\pm (q+2\nu)$ terms (\ref{qbit2}) will be minimised.

Line (\ref{line1}) is the energy gap created by the microwaves, analogous to the single-qubit case. $H_{res}$ represents numerous residual terms that contain $\nu$ and $q$ in their rotation frequencies. An expression for $H_{res}$ in the interaction picture with respect to (\ref{line1}) is provided in the appendix (\ref{right}). These terms would be expected to cancel by rotating wave arguments, however, they will be shown to contribute to two non-negligible spurious coupling effects.

Considering the effect of magnetic noise in the dressed basis, the following contribution is found:
\begin{align}
H_n = -\frac{\mu}{\sqrt{2}} \,\,\bigg( \ket{D}\bra{u} +\ket{D}\bra{d}\,+\,h.c. \bigg).
\label{considered}
\end{align}
This can be compared to (\ref{sunday}). Moving to the interaction picture with respect to (\ref{line1}) will generate shielding against magnetic noise, as has been presented before. This mechanism is maintained as one extends the discussion to multi-particle Hamiltonians.

\subsection{Two-particle Hamiltonian}

We present and simulate the entangling gate for the two-particle case, noting that a multi-particle entangling gate would also be viable. The case discussed is for the D-qubit, using the centre-of-mass mode. The breathing mode is not treated explicitly, but the effects of its presence will be discussed in Section \ref{spr}.

The single-particle Hamiltonian (\ref{line1}-\ref{hres}) needs to be re-derived for the extended $(\mathcal{H}_{ion1} \otimes \mathcal{H}_{ion2}) \otimes\mathcal{H}_{phonon}$ Hilbert space, making the necessary modifications. The term $\kappa \sigma_z (b^\dagger + b)$ in line (\ref{cc3}) enters with the same sign for each of the two qubits, provided that the centre-of-mass mode is assumed. One performs the Schrieffer--Wolff transformation of form:
\begin{align}
M\rightarrow e^{\eta (\sigma_{z1} + \sigma_{z2} ) (b^\dagger - b)}Me^{- \eta (\sigma_{z1} + \sigma_{z2}) (b^\dagger - b)}
\label{SWT2}
\end{align}
to remove the $\kappa \sigma_{zi} (b^\dagger+b)$ contributions and recover the following extra term:
\begin{align}
H_{SW2}= - \eta^2 \nu \,\,(\sigma_{z1}+\sigma_{z2} )^2. 
\label{SW2}
\end{align}
The other steps in the derivation (interaction picture, rotating wave approximation, basis change) are generalised straightforwardly to the two-qubit case to yield a generalisation of the Hamiltonian (\ref{line1}-\ref{hres}). Finally, one moves to the interaction picture with respect to the (generalised version of) microwave part (\ref{line1}) to obtain the two-qubit Hamiltonian of the final form. This step leaves the terms (\ref{qbit}-\ref{qbit2}) (in the extended Hilbert space) unaffected.

Using the definition:
\begin{align}
\sigma_+=\ket{D}\bra{0'}
\end{align}
the Jaynes-Cummings terms ((\ref{qbit}), in the extended Hilbert space) can be rewritten in the form:
\begin{align}
H_q = i\frac{\eta \Omega_g}{\sqrt{2}}\,\bigg(\sigma_{y1}+\sigma_{y2}\bigg)\bigg( e^{iqt}b - e^{-iqt}b^\dagger\bigg).
\label{JCF}
\end{align}
This expression is used to obtain the fast M{\o}lmer-S{\o}rensen gate.

The effect of the faster-oscillating terms of form (\ref{qbit2}) (in the extended space) will be minimised by parameter choice. One checks for any other unwanted interactions in the final Hamiltonian by expanding it to the second order in the Dyson series and looking for secular terms. The following additional contribution is found:
\begin{align}
\notag&H_{add}\,\,= \\
\notag\\
\notag&\,\frac{- 2 \eta^2 \nu^3}{2 \nu^2 - \Omega^2}\cdot \bigg(2\ket{DD}\bra{DD}+\ket{D0'}\bra{D0'}+\ket{0'D}\bra{0'D}\, \\
\label{rsns}&\,\, + \, \ket{DD}\bra{ud} + \ket{DD}\bra{du} + \ket{ud}\bra{DD}+ \ket{du}\bra{DD}  \bigg)
\end{align}
which affects significantly the gate performance and needs to be minimised. 

\subsection{Fast entangling gate}

Following the proposal of M{\o}lmer and S{\o}rensen \cite{sorensen2000entanglement}, a two-qubit entangling gate can be obtained from the Hamiltonian $H_q$ (\ref{JCF}). The functions $F(t)$ and $G(t)$ (defined in the M{\o}lmer-S{\o}rensen derivation) need to be set to zero, which imposes the constraint:
\begin{align}
q t = 2\pi \cdot R 
\label{cond1}
\end{align}
for integer R. Setting in addition:
\begin{align}
t \frac{\eta^2 \Omega_g^2}{q}=\frac{\pi}{4}
\label{cond2}
\end{align}
leads to the desired unitary evolution, which generates entanglement between the qubits:
\begin{align}
U_T = \text{Exp}\bigg(\frac{-i\pi}{4}\cdot (\mathbb{1}+\sigma_{y1}\sigma_{y2}) \bigg).
\label{unitary}
\end{align}

Given a value for R, the conditions (\ref{cond1}, \ref{cond2}) fix the time of the entanglement operation to: 
\begin{align}
T=\frac{\pi \sqrt{R}}{\sqrt{2}\eta \Omega_g}.
\label{gateT}
\end{align}
Furthermore, the value for $q$ is also determined:
\begin{align}
q=2 \sqrt{2R}\eta \Omega_g.
\label{q34}
\end{align}

\subsection{Minimising spurious couplings}
\label{spr}

Experimental parameters have to be chosen to minimise excitations of the other motional mode and the effect of the resonance term (\ref{rsns}). The breathing mode frequency is given by $\nu'=\sqrt{3}\nu$ (which is also the next lowest frequency in the N-particle case \cite{vsavsura2002cold}), and the introduction of the breathing mode phonon terms $\nu' b'^\dagger b'$ and $\pm \kappa' \sigma_z (b'^\dagger + b')$ in the Hamiltonian (see (\ref{cc3})) would lead to extra prefactors of form $e^{\pm \eta' (\tilde{b'}^\dagger-\tilde{b'})} $ in (\ref{sat}).

Considering the effect of such terms on the qubit-space couplings (\ref{qbit}, \ref{qbit2}), the next lowest oscillation frequency after $e^{\pm iqt}$ will be close to $e^{\pm i (\nu - \nu') t}$ (assuming $q \ll \nu$). It will be found in terms of the following functional form:
\begin{align}
\frac{\eta' \Omega_g}{\sqrt{2}} e^{ i (\nu - \nu') t} \,\, b' \approx \frac{\eta \Omega_g}{3.25} \,\, e^{ - i \cdot 0.73 \nu  t} \,\, b'
\end{align}
where we have used $\nu'=\sqrt{3}\,\nu$ and $\eta' = 3^{-3/4}\, \eta$ (see appendix \ref{etad}). This represents the effect to be minimised, which generates a contribution in the second order of the Dyson series. Comparing this coupling with the strength of the gate coupling (\ref{gateT}) leads to the condition:
\begin{align}
\notag &\frac{\eta^2 \Omega_g^2}{\nu}\ll \eta \Omega_g \\
\label{condition1} &\eta \Omega_g \ll \nu.
\end{align}
This constraint also ensures that the terms of line (\ref{qbit2}) yield a negligible effect.

Secondly, the magnitude of the terms in (\ref{rsns}) can be minimised (using the assumption $\nu^2 \gg \Omega^2$) by requiring the following:
\begin{align}
\notag &\eta^2 \nu \ll \eta \Omega_g \\
\label{condition2} & \eta \nu \ll \Omega_g.
\end{align}

Conditions (\ref{condition1}) and (\ref{condition2}), together with the expressions for $T$ and $q$ (\ref{gateT}, \ref{q34}) and the relationship $\eta \propto \nu^{-3/2}$ constrain the choice of experimental parameters and ultimately the properties of the entangling gate that can be produced within a given set of experimental limitations. We still find considerable freedom in the parameter range and choice, so that further research, especially in light of a particular experimental arrangement, would be encouraged. 

The presence of a further motional mode (or several) in the derivation would also modify the expressions for $H_{SW2}$ (\ref{SW2}) and $H_{res}$ (\ref{hres}) (in the extended space), which would mathematically alter the unwanted resonance effects to some degree. This modification, which in general would depend on the particle number, can be tackled further by analytical and numerical techniques.

There are further strategies available for the suppression of the terms (\ref{rsns}), which can be pursued. In order to suppress the first row of (\ref{rsns}), one can shift by $\eta^2\nu$ the energy level of $\ket{0'}$. This can be achieved by applying a detuned microwave field to couple $\ket{0}$ and $\ket{0'}$. We discuss this technique in appendix \ref{DSS} and also in Section \ref{MT} in the context of avoiding unwanted cross-couplings within the four-level system.

One can also suppress the leakage terms in the second row of (\ref{rsns}) by countering their first order contribution. This is done by opening an energy gap such that these transitions no longer preserve energy, thus we will remain with a higher order contribution only. Opening such an energy gap can be achieved in two ways: firstly, by introducing equal detunings $\delta_0$ to the microwave dressing fields, such that a $2\delta_0\ket{0}\bra{0}$ term is introduced in the Hamiltonian. In the dressed state basis, this additional term becomes $\delta_0\left( \ket{u}\bra{u}+\ket{d}\bra{d} \right)$, so that the energies of these levels are no longer equidistant from the dark state $\ket{D}$ (see (\ref{line1}-\ref{qbit2})). 

Secondly, one can use different Rabi frequencies for the dressing fields that operate on the different ions, such that the i'th ion is irradiated with $\Omega_i$. For this strategy only, we define $\delta_0=\Omega_1-\Omega_2$. The leakage is energetically suppressed in both strategies when ${\eta^2\nu}\ll\delta_0$, since now the second row of (\ref{rsns}) will rotate with a fast $\delta_0$ frequency.

In the numerical simulation, we pursue the first way, introducing a detuning $\delta_0$ to all the microwave dressing fields. We also implement the strategy outlined in a preceding paragraph, adding a $\ket{0} \leftrightarrow \ket{0'}$ coupling, characterised by $\Omega_z$, $\delta_z$ for each particle (see (\ref{params3})).

\subsection{Fidelity correction}

A third prominent unwanted coupling effect is found in numerical simulation and can be traced to terms in $H_{res}$, specifically, the part proportional to $\Omega_g$ (see (\ref{right})). The effect of these terms is to superimpose a fast-oscillating time dependence on some of the plots for state fidelity during the gate operation. 

The analytical treatment of this effect mirrors closely the derivation by M{\o}lmer and S{\o}rensen \cite{sorensen2000entanglement} (Section III A. Direct coupling). Firstly, we assume $\Omega \ll q+\nu$, so that the terms responsible for the disturbance can be approximated to the following expression (here quoted for the single-particle Hamiltonian):
\begin{align}
\notag H_{c:1q} = \,&\frac{\Omega_g}{2}\,\bigg( (e^{-itq-it\nu}+e^{itq+it\nu})\ket{0'}\bra{u}+\\
\notag& \,\,\,\,\,\,\,\,\,\,\,\,\,\,\,  (e^{-itq-it\nu}+e^{itq+it\nu}) \ket{0'}\bra{d} \\
\label{right2}&\,\,\,\,\,\,\,\,\,\,\,\,\,\,\,\,\,\,\,\,\,\,\,\,\,\,\,\,\,\,\,\,\,\,\,\,\,\,\,\,\,\,\,\,\,\,\,\,\,\,\,\,\,\,\,\,\,\,\,\,\,\,\,\,\,\, + h.c.\bigg).
\end{align}
Secondly, taking the desired gate evolution to be $U(t)$, one transforms the disturbance (rewritten for the two-qubit case) to the interaction picture: $H_{cI}(t)=U^\dagger(t) H_c(t) U(t)$, and considers expanding $H_{cI}(t)$ in the Dyson series to evaluate the magnitude of the disturbance.

Two simplifying approximations are made. Firstly, $U(t)$ is taken to be slowly-varying in comparison to $H_{c}(t)$, so that it can be regarded as constant when performing the Dyson series integrals. Secondly $H_{cI}(t)$ is evaluated in the vicinity of the endpoint of the gate operation $(t=T)$, where $U(t)$ takes a simple form (\ref{unitary}) and is approximated to be time-independent.

Obtaining an expression for $H_{cI}(t)$ in such a manner to the second order in the Dyson series, one can calculate the fidelity of certain output states, given a particular input state. One also needs to account for the fact that an interaction picture has been adopted. Again, we use a different definition of fidelity to the paper by M{\o}lmer and S{\o}rensen: $F(\ket{\Psi_c}, \rho) = \sqrt{\bra{\Psi_c}\rho\ket{\Psi_c}}$, so that state probabilities are given by $F^2$.

\begin{figure}
\includegraphics[width=0.5\textwidth]{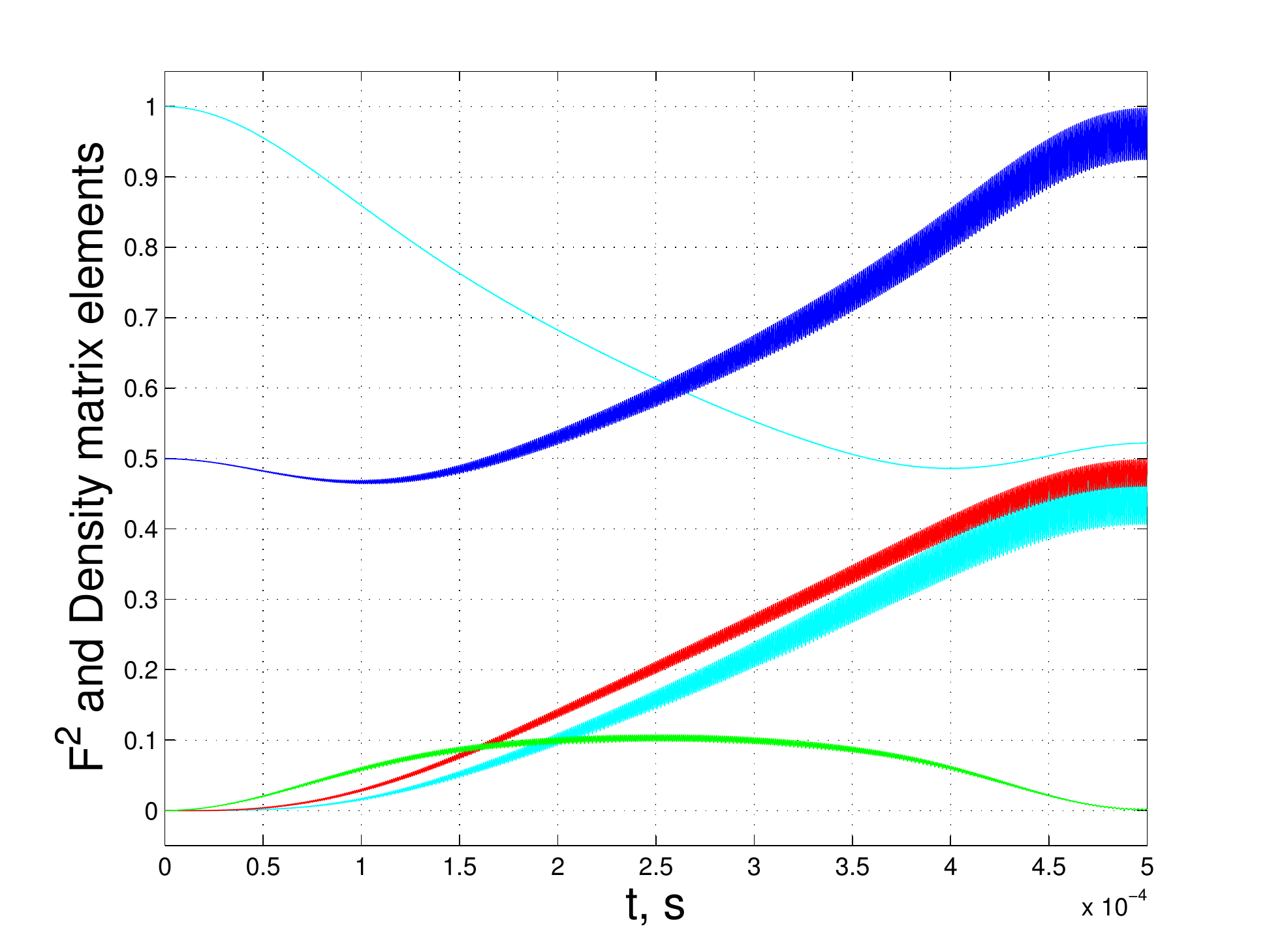}
\includegraphics[width=0.5\textwidth]{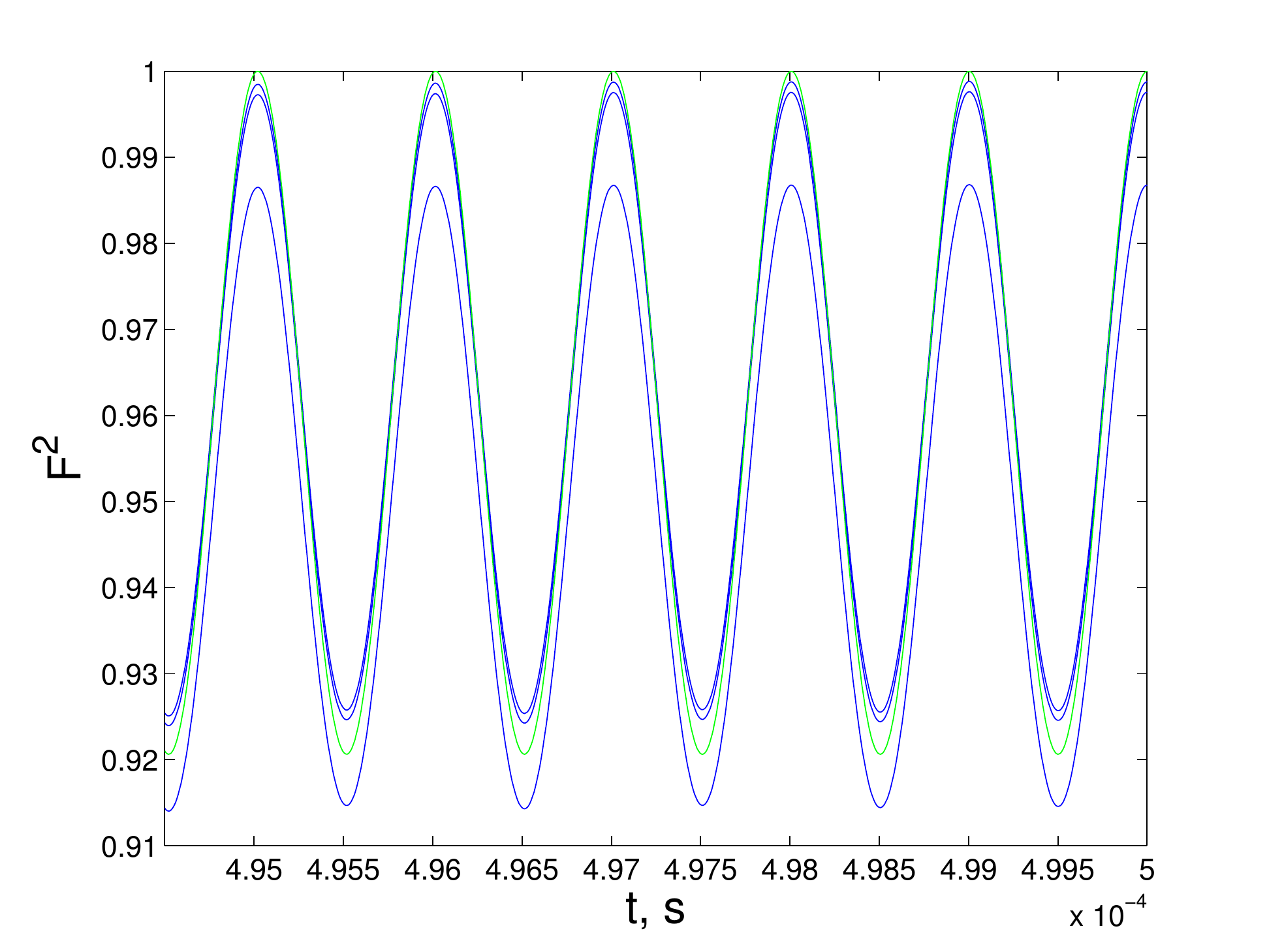}
\caption{\underline{TOP}: Squared fidelity and other density matrix elements for the two-qubit entangling gate. An input state of $\ket{DD}$ is used and the simulation parameters are specified in (\ref{params3}), except that the heating rate of 10 phonons/s only has been plotted. The first curve (counting from above at $t\approx 0.25$ ms) represents the squared fidelity of $\ket{DD}$ (light blue), where no oscillatory component is found. The second curve (blue) is the squared fidelity of $\frac{1}{\sqrt{2}}(\ket{DD}+i\ket{0'0'})$. The third (red) is the imaginary part of $\rho_{\ket{DD},\ket{0'0'}}$, the fourth (light blue) is the squared fidelity of $\ket{0'0'}$, and the last curve (green) is the real part of $\rho_{\ket{DD},\ket{0'0'}}$. \underline{BOTTOM}: A magnified segment of the squared fidelity plot for $\frac{1}{\sqrt{2}}(\ket{DD}+i\ket{0'0'})$ during the gate operation. The result of the calculation (\ref{corr}) is plotted in green. The results for the three different phonon heating rates (\ref{params3}) are plotted in blue.}
\label{2qplots}
\end{figure}

Beginning in the state $\ket{DD}$ and calculating the fidelity of $\ket{DD}$ at the end of the gate operation, one recovers $F^2= \frac{1}{2}$. This is consistent with the unitary gate evolution (\ref{unitary}) and is verified in the numerical simulation (see Figure \ref{2qplots} top), where no oscillatory effect is observed. In contrast, starting in the state $\ket{DD}$ and calculating the fidelity of $\frac{1}{\sqrt{2}}(\ket{DD}+i\ket{0'0'})$, the following is obtained:
\begin{align}
F^2=1 - \frac{2\Omega_g^2}{(q+\nu)^2}(\sin{(q+\nu)t})^2+\mathcal{O}\bigg(\frac{\Omega_g^4}{(q+\nu)^4}\bigg).
\label{corr}
\end{align}
An oscillatory correction is thus introduced to the fidelity of the entanglement operation. Numerical simulation suggests that (\ref{corr}) predicts very accurately the frequency and the amplitude of the oscillations observed (see Figure \ref{2qplots} bottom). A simliar calculation can be carried out for any other input and comparison states.

This oscillatory effect can be minimised by reducing $\Omega_g/\nu$, or by adjusting precisely the gate duration. Higher trap frequency $\nu$ will lead to a greater accuracy requirement for the length of the gate pulse. One could also use pulse shaping techniques in order to increase the timing accuracy \cite{roos2008ion, hayes2012coherent}.

In the absence of any such mitigating techniques being employed, we have presented the mathematics for deriving the oscillation parameters analytically for arbitrary input and output states. The process fidelities can be computed using these techniques.

\subsection{Simulation}
\label{2qsim}
Numerical simulation of the two-qubit entangling gate is carried out to demonstrate its feasibility. We simulate a Hamiltonian of the form (\ref{thor}-\ref{cc3}), extended to the two-qubit case. A single motional mode is used: the centre-of-mass mode. The effects of magnetic noise in the multi-qubit case have been shown to be directly analogous to the single-qubit arrangement (see (\ref{considered})), where sufficiently strong microwave dressing field renders the disturbance negligible. No magnetic noise or any other random noise effects have been included in the present simulation.

The following parameters are used:
\begin{align}
&\notag \Omega= 2\pi \cdot 20\,\text{kHz} \\
&\notag \delta_0= 2\pi \cdot 2\,\text{kHz} \\
&\notag\Omega_g=2\pi \cdot 100\,\text{kHz} \\
&\notag\eta = 0.0071 \\
&\notag\nu = 2\pi \cdot 500 \,\text{kHz} \\
&\notag \Omega_z= 2\pi \cdot 10\,\text{kHz} \\
&\notag \delta_z= 2\pi \cdot 1\,\text{MHz} \\
&\notag n=0 \\
& \notag R = 1 \\
& \text{Heating rate: 0, 10, 100 phonons/s}
\label{params3}
\end{align}
In this simulation, we have included the additional detuned $\ket{0} \leftrightarrow \ket{0'}$ coupling, characterised by $\Omega_z, \delta_z$ for each of the two particles. We have also detuned the two microwave dressing fields (Rabi frequency $\Omega$) of each trapped particle by $\delta_0$. These steps are taken as strategies for minimising the effects of (\ref{rsns}) on gate fidelity, as discussed at the end of Section \ref{spr}.

This parameter choice yields the gate time $T=0.5$ ms, and sideband detuning $q = 2\pi \cdot 2$ kHz. The constant of proportionality linking $\nu$ and $\eta$ (see appendix \ref{etad}) is obtained for the ${}^{171}Yb^+$ ion and magnetic gradient of $46$ T/m. Current laboratory technology has enabled gradients of up to $24$ T/m to be realised \cite{Hensn}, so that our parameter choice is not unrealistic. Moreover, in the macroscopic ion traps the magnetic gradient is created by two anti-Helmholtz coils, which are placed far away from the trap, resulting in limited gradient \cite{wund, Hensn}. However, in future planar traps, the gradient can be very high due to surface proximity. This will have the effect of also increasing the heating rate, though experimental techniques for reducing the heating effects could be implemented to remedy the problem \cite{kunert2014planar}.

Figure \ref{2qplots} plots squared state fidelities and density matrix elements for the duration of the gate operation. An input state of $\ket{DD}$ has been used, so that the worst-case scenario has been taken: considering (\ref{rsns}), it is seen that $\ket{DD}$ is affected the most by the unwanted resonance effect. Hence, if one wanted to obtain the process fidelity for the operation, which amounts to averaging over the input states, then a higher maximal fidelity would be reached.

The Figure gives clear evidence for the feasibility of the entangling gate. Also, the oscillatory correction to the fidelity of the target state $\frac{1}{\sqrt{2}}(\ket{DD}+i\ket{0'0'})$ is found to be in good agreement with the mathematical description (\ref{corr}), as seen in Figure \ref{2qplots} bottom panel. The gap in the heights of the oscillatory plots here can be explained by the presence of further noise effects in the simulation and the different heating rates used.

Using this set of parameters yields gate fidelities of $F^2=99.88\%$, $F^2=99.76\%$, and $F^2=98.68\%$ for heating rates of $0, 10, 100$ phonons/s, respectively. These values are attained precisely at the end of the gate operation $(t=T)$ and therefore would assume a near-perfect control of the experimental execution of the gate. It is acknowledged that the oscillatory effect, as plotted in Figure \ref{2qplots} bottom panel, introduces a deviation of amplitude $\Delta (F^2) \approx 8\%$ and period close to $ 1 \mu s$ into the fidelity plots, making the attainable gate fidelity very sensitive to precise experimental execution. Small deviations in the gate duration or possible drifts in other experimental parameters could thus have a strong detrimental effect on the fidelity attained. This represents an unavoidable source of noise and a challenge to be tackled in future experimental work.

The small reduction in attainable gate fidelity with increased heating rate provides evidence for good robustness of the scheme against heating, as is also the case for the original M{\o}lmer-S{\o}rensen design. There is considerable scope for variability in the values of the experimental parameters used for the gate. It is generally found that having access to higher magnetic gradient enables the design of an entangling gate with better properties.

\section{Beyond the Linear regime}
\label{F18}

This Section discusses extensions and generalisations of the dressed-state approach to the regime where non-linear Zeeman shift plays a prominent role. The case of ${}^{171}Yb^+$ is discussed in particular. We delineate precisely the 'Linear' regime for this physical system, which is the region of validity for the derivations presented above. We also define and discuss a 'Non-linear' regime, exemplified by the recent work of Webster et al. \cite{webster2013simple}. The relative merits of these two parameter ranges are then considered, together with a possible strategy for attaining either experimentally by means of microwave dressing fields.

\subsection{Hyperfine Zeeman shift in ${}^{171}Yb^+$} 

The four-level system depicted in Figure \ref{f1} can be realised using the $F=\{0, 1\}$ hyperfine ground state of ${}^{171}Yb^+$ with non-zero external magnetic field. The $\ket{1}$ and $\ket{-1}$ states would correspond to the $m_f = \pm 1$ levels of the $F=1$ triplet, $F =1, m_f = 0$ level would yield the $\ket{0'}$ state and $\ket{0}$ would be represented by the singlet $F=0$ state. The study by Blatt et al. \cite{blatt1983precise} presents a detailed energy-level diagram of the system as well as provides an accurate measurement of the singlet-triplet energy splitting, which is approximately $A = 2\pi \cdot 12.6$ GHz.

The $\ket{\pm 1}$ states respond exactly linearly to external magnetic field B, with a change in energy of $\pm \mu_B B$. The response of $\ket{0'}$ and $\ket{0}$ can be approximated to the lowest order by $\pm (\mu_B B)^2/A$ \cite{foot2004atomic}. For any non-zero external field, there is therefore an inevitable discrepancy between the $\ket{-1} \leftrightarrow \ket{0'}$ and $\ket{0'} \leftrightarrow \ket{1}$ resonant frequencies, which can be well approximated by the (positive) figure:
\begin{align}
\Delta = \frac{2(\mu_B B)^2}{A}. 
\label{Delta}
\end{align}
This enables the explicit definition of two simplified physical regimes.

\subsection{Linear regime} 

The gates presented in the previous Sections are built on the assumption of negligible $\Delta $, so that addressing of both $\ket{-1} \leftrightarrow \ket{0'}$ and $\ket{0'} \leftrightarrow \ket{1}$ pairs can be achieved by the same $\Omega_g$ field. Addressing one pair of levels exactly on resonance would mean that the other pair is addressed with the (positive) detuning equal to $\Delta$. It is necessary to preserve this second coupling as a desired effect, with the contribution due to $\Delta$ being negligible.

In the single-qubit case, considering the Rabi model \cite{gerry2005introductory}, making the two interactions equivalent would require:
\begin{align}
&\notag \Omega_g \approx \sqrt{\Omega_g^2+\Delta^2} \\
& \Omega_g^2 \gg \Delta^2.
\label{linear}
\end{align}
In the multi-qubit case, where the gate interaction strength is of the order $\eta \Omega_g$, one requires $\Delta$ to obey the following constraint:
\begin{align}
\eta \Omega_g \gg \Delta.
\label{linearxx}
\end{align}
In both cases, an upper limit on the permissible magnetic field is placed by the strength of the RF fields employed.

In the Sections above, we have also assumed that magnetic noise affects prominently the $\{\ket{-1}, \ket{1}\}$ states, but negligibly the $\{\ket{0}, \ket{0'}\}$ pair of levels. This relies on the assumption of small magnetic field. Comparing the sensitivity of $\ket{\pm 1}$ to magnetic noise with the (B-field dependent) sensitivity of $\ket{0'}$ leads to the requirement:
\begin{align}
B \ll 0.45 \,\, \text{T}.
\label{cond}
\end{align}

Raising the magnetic field beyond this value will introduce noise effects into the system not corrected for. In conjunction with the magnetic gradient used (\ref{params3}), this consideration leads to an upper limit on the permissible axial range of the experimental configuration, confining the ion arrangement to the size of $\ll 1$ cm on present numbers. This constraint will be satisfied in the case of a simple ion chain. However, it may become problematic in the case of a more elaborate design with several gate regions to obey the size requirement, with the resulting problem of additional noise sources to be considered.

\subsection{Non-linear regime} 

This regime is defined as the instance when both $\ket{-1} \leftrightarrow \ket{0'}$ and $\ket{0'} \leftrightarrow \ket{1}$ pairs can be unambiguously individually addressed, without affecting the other coupling. In this case, the coupling of the other pair, with the detuning equal to $\Delta$, would represent an unwanted effect to be made negligible. This is the case for prominent $\Delta$, such that the Stark shift approximation \cite{james2007effective} applies. The condition is: 
\begin{align}
& \Omega_g \ll \Delta
\label{nonlinear}
\end{align}
which also ensures that the magnitude of the energy shift of $\ket{0'}$, $\Omega_g^2/4\Delta$, is small compared to its Zeeman response, $\Delta/2$, and therefore amounts to a negligible effect.

Experiments within the non-linear regime have been conducted by Webster et al. \cite{webster2013simple}, also citing the condition (\ref{nonlinear}). A field of $9.8$ G is used to generate a measured frequency discrepancy $\Delta = 2 \pi \cdot 29(1)$ kHz in agreement with (\ref{Delta}). Radio frequency fields of strength $\Omega_g = 2\pi \cdot 1.9$ kHz have been employed.

The authors have discussed how the non-linear regime enables the realisation of arbitrary single-qubit $\sigma_\phi$ gates using a single radio frequency field. Also, the authors note that a $\sigma_z$ gate could be realised by the use of a single detuned radio field.
      
The facility of individual addressing does offer clear experimental advantages, however, it may also be the case that greater sensitivity to magnetic noise is introduced as well. Considering the criteria (\ref{linear}, \ref{nonlinear}), it is probable that the non-linear regime will involve stronger B-fields than the linear regime, particularly for the arrangement of an ion chain. If the condition (\ref{cond}) is broken, this would introduce non-negligible noise in the energy of $\ket{0'}$, which is not shielded against in the present set-up. 

A further problem for the non-linear arrangement might arise in the attainment of individual addressing in an ion chain, due to the significantly non-linear dependence of the energy spacings for individual qubits.

\subsubsection{Single-qubit gates} 
\label{see}
A variety of ways to realise universal single-qubit rotations is possible in the non-linear regime. In addition to the proposals by Webster et al. \cite{webster2013simple}, it is noted that individual addressing ($\phi_- \neq \phi_+$) allows for the basic gate arrangement (Section \ref{basic}) to yield both the $\sigma_x$ and the $\sigma_y$ gates for the B and D-qubits. An extra error source to consider would be the instability of the radio frequency fields ($ \delta_{\Omega_g} = \Omega_{g-}-\Omega_{g+}$), due to two fields being necessary.

No extra effort would be required to realise adiabatic transfer, and the adiabatic $\sigma_z$ gate (Section \ref{adiz}) would be realisable by the usage of two RF fields per trapped particle. Further, the two $\sigma_z$ gates presented in the appendix are also a feasible alternative. In every case where two RF fields are being used, the small extra noise contribution due to $\delta_{\Omega_g}$ would need to be considered. 

\subsubsection{Multi-qubit gate} 

The linear response of $\ket{-1}$ and $\ket{1}$ to magnetic field in the Ytterbium system permits the realisation of magnetic-gradient-induced coupling for any strength of the B-field, which is a crucial ingredient for the entangling gate. The reproduction of the M{\o}lmer-S{\o}rensen gate presented in this paper (Section \ref{b52}) would be possible in the non-linear regime by the usage of four radio frequency fields per trapped particle. 

Separate coupling of the magnetic-sensitive states is found to offer no clear mathematical advantage in the construction of the entangling gate. It is possible to employ two radio frequency fields (in two arrangements) and reach an entangling Hamiltonian of form similar to (\ref{line1}-\ref{hres}). However, the speed of the resultant gate is reduced by $1/2$. 

Moreover, it is the property of the linear regime multi-qubit gate that the zeroth order in $\eta$ is canceled within the qubit space, in the dressed basis, leaving only terms to the first order in $\eta$ (see (\ref{qbit}-\ref{qbit2})). This property ceases to hold for a gate that is built using two RF couplings per trapped particle. As a result, unwanted zeroth order terms of form $\Omega_g e^{\pm i (q+\nu) t}$ are introduced within the qubit space. This would lead to a more demanding set of constraints on the gate parameters. 

These considerations make the M{\o}lmer-S{\o}rensen gate harder to realise in the non-linear regime. 

\subsection{Mediating technique} 
\label{MT}

The linear and non-linear regimes are compounded by an intermediate region where neither perfect individual nor perfect mutual addressing in the qubit space are possible. The facility to reach either regime can be hampered by the existence of an upper limit on the B-field strength (\ref{cond}), as well as experimental limitations on the gate time or $\Omega_g$. In such cases, an intermediate regime may be inevitable, with the ensuing presence of spurious couplings within the qubit space.

As an alternative to tackling explicitly such couplings, the technique of dressed Stark shift (appendix \ref{DSS}) offers a way of tuning $\Delta$ by means of microwave fields. Such a process would potentially provide easy mediation between the linear and non-linear regimes. Using a detuned microwave field specified by $\Omega_z, \delta_z$ to induce a $\ket{0} \leftrightarrow \ket{0'}$ coupling, together with the two microwave dressing fields, leads to the following additional term in $\Delta$:
\begin{align}
\Delta = \frac{2(\mu_B B)^2}{A}+\frac{\delta_z \Omega_z^2}{ \Omega^2 - 2 \delta_z^2}
\label{Delta2}
\end{align}
subject to the conditions for fast oscillation (\ref{SSc}). This suggests the possibility of tuning $\Delta$ with the help of a second physical process. The above result is found by considering $\Omega_z$ and two microwave dressing fields only, so the potential cross-couplings due to the presence of RF fields would also need to be examined.

Within an ion chain, it is likely that a single $\Omega_z$ field would generate couplings between the $\ket{0}$ and $\ket{0'}$ states of all the ions involved, so that no individual control over $\delta_z$ and $\Omega_z$ would be attainable. However, independent tuning of $\Delta$ would still be possible, in principle, by means of the $\Omega$ dressing fields, which are well separated in frequency space. 

Provided that the tuning of $\Delta$ can be realised with attainable experimental parameters, dressed Stark shift offers a way of realising both linear and non-linear regimes using modest magnetic field strength. This would be of advantage for both single and multi-qubit designs.

\section{Prospects for radio-wave-only quantum gates}

Section \ref{basic} and the corresponding simulations (Section \ref{numB}) have illustrated how a single-qubit gate of working time in the range of ms can be realised using RF fields of strength $2\pi\cdot177$ Hz, and relying on the microwave dressing to provide the magnetic shielding effect. 

One notes, in addition, that a scheme would also be possible, where the radio frequency fields both generate the gate coupling and provide magnetic shielding via the introduction of a time-dependent phase to the magnetic noise terms. Considering the D-qubit case (\ref{h15}) with only the $\Omega_g$ part present, it is clear that rotations will be introduced to the magnetic noise term ((\ref{sunday}), setting $\delta_\Omega = 0$). A separate Dyson series analysis needs to be carried out to evaluate exactly the noise terms. One finds that, for such a set-up, noise suppression would occur for $\Omega_g \gg SD_\mu$. It is found numerically that shielding is indeed accomplished for sufficiently high $\Omega_g$, yielding robust gates on the timescale of $\mu$s. In the case of the non-linear regime, this arrangement would indeed permit the realisation of universal single-qubit rotation using radio-wave addressing only (see Section \ref{see}). 

It is an interesting research venue to pursue whether a feasible radio-wave-only entangling gate could also be designed. In the absence of a viable shielding mechanism being generated by the radio frequency fields without the microwaves, it may be possible to out-pace the noise effects by realising a gate of sufficiently high speed. Also in the microwave-dressed state approach, the possibility of realising an entangling gate that out-paces the noise effects would be worthy of further study.

\section{Conclusion}

We have demonstrated the feasibility of universal quantum computing using microwave-dressed states in trapped ions or any other suitable system where sufficient coupling between atomic and motional states can be obtained. Both single and multi-qubit quantum operations have been proposed and their resilience against noise sources analysed in detail. This raises the prospects of microwave/radio wave-driven quantum computation as an exciting venue for future research. An interesting question to address would be the implementability of other multi-qubit gate designs in the dressed-state system to compete with the M{\o}lmer-S{\o}rensen design.

\section*{Acknowledgement}

We gratefully acknowledge the support of Shai Machnes in the computational aspects of the project. We also thank Michael Drewsen for helpful discussion. AR thanks Winfried Hensinger and Seb Weidt for useful discussions and acknowledges the support of the European Commission (STREP EQuaM), the Ministry of Science and Culture in Lower Saxony, and the Israeli Science foundation. MBP was supported by the EU Integrating Project SIQS, the EU STREP EQuaM and an Alexander von Humboldt Professorship.

\appendix

\section{$\sigma_z$ gate via dressed $\ket{0}\leftrightarrow \ket{0'}$ Stark shift}
\label{DSS}

Detuned $\ket{0} \leftrightarrow \ket{0'}$ coupling enables a phase to be induced in $\ket{0'}$ and a $\sigma_z$ gate to be realised using the effect of Stark shift. It is shown how microwave dressing can be added to the process to protect it from noise effects.

Removing the radio frequency fields in (\ref{ham}) and setting for the D-qubit case:
\begin{align}
\notag &\theta_z = 0\\
\notag &\Omega_{+/-} = \Omega\\
 &\theta_-=\theta_+ = 0
\end{align}
one moves to the dressed basis to obtain:
\begin{align}
H=\notag&\,\frac{\Omega}{\sqrt{2}}\, \bigg(\ket{u}\bra{u} - \ket{d}\bra{d}\bigg)\,+ \\
\label{h19}\,&\,\frac{\Omega_z}{2\sqrt{2}}\, \bigg(\ket{u}\bra{0'}e^{it\delta_z} -\ket{d}\bra{0'}e^{it\delta_z}\bigg)\,+\,h.c.
\end{align}
together with a noise contribution of form (\ref{sunday}).

Moving to the interaction picture with respect to the $\Omega$-term will cause the noise terms to rotate. In addition, the time-dependence in the term proportional to $\Omega_z$ will be modified. It still creates a Stark-shift-like effect, modifying the energies of $\ket{u}$, $\ket{d}$ and $\ket{0'}$ in the second order of the Dyson expansion. In particular, the addition to the qubit space takes the form:
\begin{align}
H_{add}=\,&\frac{\delta_z \Omega_z^2}{2 \Omega^2 - 4 \delta_z^2}\,\ket{0'}\bra{0'}
\end{align}
and enables a $\sigma_z$ gate to be realised.

The two Stark-shift-like processes in the above derivation rely on the following experimental constraint:
\begin{align}
|\Omega \pm \sqrt{2}\delta_z| \gg \Omega_z.
\label{SSc}
\end{align}

\section{$\sigma_z$ gate via radio wave Stark shift}

\label{RSS}
A $\sigma_z$ gate is presented that requires independent addressing of the magnetic levels by separate radio frequency fields. The D-qubit case is shown. Canceling the $\ket{0} \leftrightarrow \ket{0'}$ coupling and setting in (\ref{ham}):
\begin{align}
\notag & \Omega_{+/-} = \Omega\\
\notag &\theta_-=\,\theta_+=0\\
\notag &\phi_+=\pi\,\,\,,\,\,\phi_-=0\\
&\delta_+=-\delta\,\,\,,\,\,\,\delta_-=\delta
\end{align}
one recovers
\begin{align}
H=\notag&\,\frac{\Omega }{\sqrt{2}} \, \bigg(\ket{u}\bra{u} - \ket{d}\bra{d}\bigg)\,+ \\
\,&\,\frac{ \Omega_g}{\sqrt{2}}\, \bigg(\ket{0'}\bra{D}e^{i \delta t} + h.c. \bigg).
\end{align}
The radio wave part yields the $\sigma_z$ gate between $\ket{D}$ and $\ket{0'}$, using the standard Stark shift approximation \cite{james2007effective}:
\begin{align}
H_{rw} \approx  \frac{ \Omega_g^2}{2 \delta}\bigg(\ket{0'}\bra{0'}-\ket{D}\bra{D}\bigg).
\end{align}
The condition of validity for the last step is:
\begin{align}
\Omega_g \ll \delta.
\end{align}

\section{Effective Lamb-Dicke parameter}
\label{etad}

Section \ref{b52} makes use of the following definitions for $\kappa$ and $\eta$:
\begin{align}
\kappa = \partial_z(\beta(z)) \frac{ \zeta}{\sqrt{2m \nu}}
\label{kappa}
\end{align}
where $\beta(z)$ is one half of the energy spacing between the magnetic-sensitive levels (Figure \ref{f4}), $m$ is the ion mass, $z$ is the axial trap co-ordinate and $\nu$ is the frequency of the motional mode employed. $\zeta$ is a translation factor equal to $1/\sqrt{2}$ for the two-particle case (both for the centre-of-mass and the breathing modes). For the N-particle case, $\zeta=1/\sqrt{N}$ for the centre-of-mass mode and $\zeta$ is close to $1/\sqrt{N}$ for the other modes \cite{vsavsura2002cold}.

We also define:
\begin{align}
\eta = \kappa/\nu.
\end{align}
In the case where interactions are created by microwave/radio wave light and the conventional Lamb-Dicke parameter is essentially zero, $\eta$ can be thought of as the effective Lamb-Dicke parameter. For clarity of presentation, this definition differs by a factor $1/2$ from the treatment by Mintert et al. \cite{mintert2001ion}.

\section{Explicit expression for $H_{res}$}
In the interaction picture with respect to the microwave part (\ref{line1}), the residual Hamiltonian in (\ref{hres}) reads:
\begin{align}
\notag H_{res} = \,&\frac{1}{2}\,\bigg(\eta \Omega(-e^{it\nu - it\Omega/\sqrt{2}} b^\dagger + e^{-it\nu-it\Omega/\sqrt{2}} b)\ket{D}\bra{u}+\\
\notag&\,\,\,\,\, \Omega_g (e^{-itq-it\nu-it\Omega/\sqrt{2}}+e^{itq+it\nu-it\Omega/\sqrt{2}})\ket{0'}\bra{u}+\\
\notag&\,\,\,\,\, \eta \Omega(e^{it\nu+it\Omega/\sqrt{2}} b^\dagger-e^{-it\nu+it\Omega/\sqrt{2}} b) \ket{D}\bra{d}+\\
\notag& \,\,\,\,\, \Omega_g (e^{-itq-it\nu+it\Omega/\sqrt{2}}+e^{itq+it\nu+it\Omega/\sqrt{2}}) \ket{0'}\bra{d} \\
\label{right}&\,\,\,\,\,\,\,\,\,\,\,\,\,\,\,\,\,\,\,\,\,\,\,\,\,\,\,\,\,\,\,\,\,\,\,\,\,\,\,\,\,\,\,\,\,\,\,\,\,\,\,\,\,\,\,\,\,\,\,\,\,\,\,\,\,\, + h.c.\bigg).
\end{align}

\bibliography{references}

\end{document}